\newif\ifcheckpagelimits
\checkpagelimitstrue
\checkpagelimitsfalse

\ifcheckpagelimits
 \documentclass[nofootinbib,prl,aps,twocolumn,showpacs,showkeys,%
 amsmath,amssymb,superscriptaddress,final,reprint,floatfix,longbibliography]{revtex4-1}
 \newcommand{\todo}[1]{}
\else
 \documentclass[pra,aps,twocolumn,showpacs,showkeys,%
 amsmath,amssymb,superscriptaddress,final,reprint,floatfix,longbibliography]{revtex4-1}
 \newcommand{\todo}[1]{{\pdfmargincomment[icon=Note,color=pink]{#1}}}
\fi

\usepackage{lineno}
  \usepackage{mathptmx}
\usepackage{subfigure}
\usepackage{dcolumn}
\usepackage{amsmath,amssymb}
\usepackage{bm}
\usepackage{color}
\usepackage{overpic}
\usepackage{latexsym}
\usepackage{epstopdf}
\usepackage{color}
\usepackage[english]{babel}
\usepackage{latexsym}
\usepackage{stmaryrd}

\usepackage{braket}

\definecolor{mygrey}{gray}{0.35}
\definecolor{myblue}{rgb}{0.2,0.2,0.8}
\definecolor{myzard}{cmyk}{0,0,0.05,0}
\definecolor{mywhite}{rgb}{1,1,1}
\definecolor{myred}{rgb}{1,0.,0.3}

\usepackage[colorlinks=true,citecolor=myblue,linkcolor=myred]{hyperref}

 \def\ee{\mathord{\rm e}}
 
 \def\ii{\mathord{\rm i}}

\def\half{\textstyle\frac{1}{2}}

\renewcommand{\ii}{{\rm i}}
\renewcommand{\ee}{{\rm e}}

\begin{document}

\title{Hybrid quantum magnetism in circuit-QED: from spin-photon waves to many-body spectroscopy}

\author{Andreas Kurcz}
\affiliation{Instituto de F\'{\i}sica Fundamental, IFF-CSIC, Calle Serrano 113b, Madrid E-28006, Spain}
\author{Alejandro Bermudez}
\email[Email to: ]{bermudez.carballo@gmail.com}
\affiliation{Instituto de F\'{\i}sica Fundamental, IFF-CSIC, Calle Serrano 113b, Madrid E-28006, Spain}
\author{Juan Jos\'e Garc\'{\i}a-Ripoll}
\affiliation{Instituto de F\'{\i}sica Fundamental, IFF-CSIC, Calle Serrano 113b, Madrid E-28006, Spain}

\begin{abstract}
We introduce a model of quantum magnetism induced by the non-perturbative exchange of microwave photons between distant superconducting qubits. By interconnecting qubits and cavities, we obtain a spin-boson lattice model that exhibits a quantum phase transition where both qubits and cavities spontaneously polarize. We present a many-body ansatz that captures this phenomenon all the way, from a the perturbative dispersive regime where photons can be traced out, to the non-perturbative ultra-strong coupling regime where photons must be treated on the same footing as qubits. Our ansatz also reproduces the low-energy excitations, which are described by hybridized spin-photon quasiparticles, and can be probed spectroscopically from transmission experiments in circuit-QED, as shown by simulating a possible experiment by Matrix-Product-State methods.
\end{abstract}

\pacs{TBD}
\ifcheckpagelimits\else
\maketitle
\fi

Quantum magnetism and low-dimensional spin models represent a cornerstone in the foundations of many-body physics\ \cite{auerbach}. The existence of integrable models\ \cite{Heisenberg_bethe, Ising_XY}, exact diagonalization\ \cite{exact_diagonalization} combined with finite-size scaling\ \cite{fss}, or numerical methods such as Density Matrix Renormalization Group\ \cite{dmrg}, proved essential for the development of new concepts and theories, e.g. quantum phase transitions\ \cite{review_qpt} and thermalization\ \cite{review_thermalization}. Despite this success, the study of non-equilibrium magnetism is generally hampered by the lack of exact solutions and efficient numerical methods. Moreover, even at equilibrium, as we abandon 1D or quasi-1D problems, frustration and disorder turn these models into NP-complete problems.

An alternative approach is to study experimentally a particular Hamiltonian, implementing it in highly controllable quantum-optical setups, the so-called \textit{quantum simulation} paradigm\ \cite{QS}. Nowadays, quantum simulation of spin models typically relies on perturbative processes, such as the exchange of phonons in ion crystals\ \cite{porras_ising,1d_Ising_ions, 2d_Ising_ions} or the super-exchange of atoms in optical lattices\ \cite{super-exchange,bloch_spin_waves,esslinger_magnetism}. These mechanisms constrain the spin-spin interaction strength, posing serious technological challenges to overcome noise\ \cite{porras_ising,1d_Ising_ions, 2d_Ising_ions} and thermal effects\ \cite{super-exchange,bloch_spin_waves,esslinger_magnetism}. It would be thus highly desirable to move away from the perturbative regime. In Ref.\ \cite{greiner_ising}, this is achieved by mapping the Ising model to a tilted Bose-Hubbard model, converting the boson tunneling amplitude into the relevant scale for interactions. However, this mapping is only valid for a certain region of the phase diagram.

\begin{figure}[b]
 \ifcheckpagelimits\else%
 \centering
 \includegraphics[width=1\linewidth]{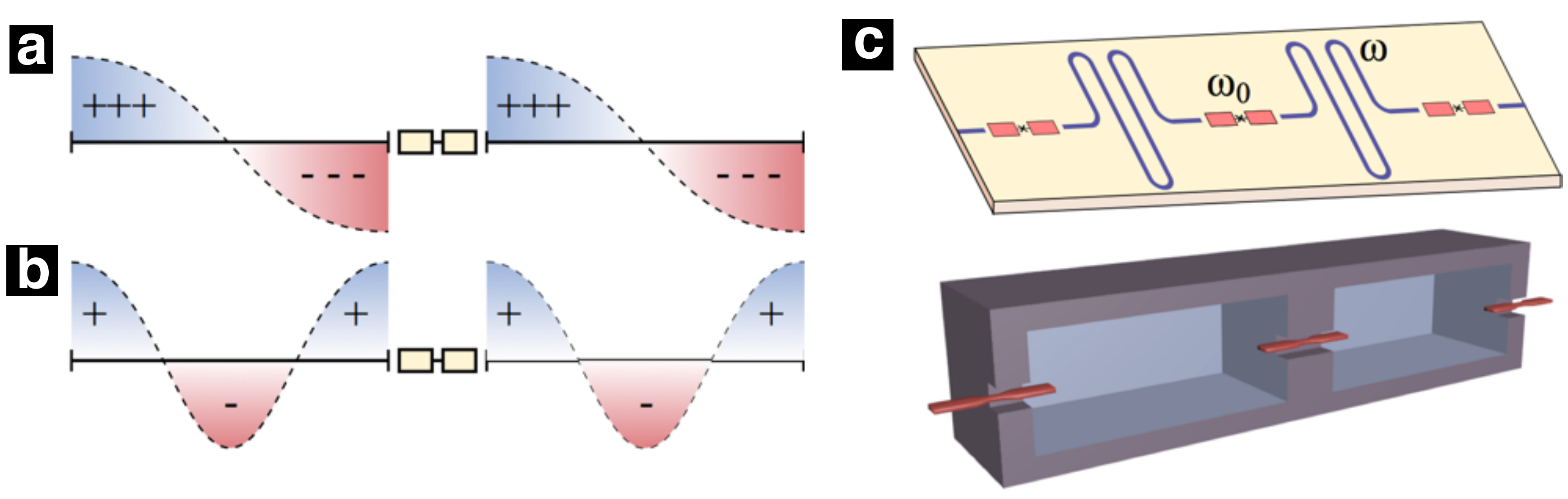}
 \fi
 \caption{{\bf Resonators coupled by superconducting qubits}. The coupling may be (a) ferromagnetic $ H_{\rm sb} = \sum_{i} g\sigma_i^x (a_i + a_{i+1}+{\rm H.c.}),$ or (b) antiferromagnetic $ H_{\rm sb} = \sum_{i} g\sigma_i^x (a_i - a_{i+1}+{\rm H.c.})$, depending on the arrangement of the cavities or the position of the qubit relative to the modes. (c) Two possible setups based on coplanar microwave guides or three-dimensional superconducting cavities which are connected through high-quality transmons.}
 \label{fig:setup}
\end{figure}

In this work, we aim at reproducing the full phase diagram and the complete dispersion relation of a critical model in the Ising universality class. To do so, we study a different type of models where the quantum magnetism arises non-perturbatively from the strong coupling between spins and bosons, $H =H_{\rm s}+H_{\rm b}+H_{\rm sb}$, where
\ifcheckpagelimits\else%
\begin{equation}
 H = \frac{\omega_0}{2}\sum_i \sigma^z_i + \sum_r \omega a^\dagger_r a_r
 + \sum_{ir} g_{ir} \sigma_i^x (a^\dagger_r + a_r).
 \label{eq:model}
\end{equation}%
\fi%
Here, the spins (bosons) are represented by Pauli matrices $\sigma_i^x,\sigma_i^y,\sigma_i^z$ (creation-annihilation operators $a_r^{\dagger},a_r$), and we have introduced the spin (boson) resonance frequency $\omega_0 (\omega)$, and the spin-boson coupling strength $g_{ir}$.
We will show that {\it (i)} these models can be implemented using standard superconducting qubits and resonators for the spins and bosons, respectively [Fig.\ \ref{fig:setup}]; {\it (ii)} the models exhibit an Ising-type quantum phase transition ---(anti)ferromagnet to paramagnet---, where both qubits and resonators spontaneously polarize as a function of  $\omega_0/\omega$ and $g/\omega$; {\it (iii)} the nature of the transition persists all the way, from a regime of weak spin-spin couplings $|J|\ll |g|\ll\omega$ to that of strong and ultra-strong quantum magnetism $|J|\lesssim |g|\approx\omega,\omega_0$ leading to high qubit-photon  entanglement.

The second goal of this work is to develop observational techniques to probe the many-body properties of these quantum magnets. These tools revolve around the idea of {many-body quantum spectroscopy}: the  system is probed with propagating fields that excite quasiparticles, which can then be measured in transmission or emission experiments. Using a many-body ansatz, Green's function techniques, and state-of-the-art Matrix Product State (MPS) simulations, we will show how to fully reconstruct the quasiparticle spectrum from such measurements, probing the static and dynamical critical exponents, and opening the door to further nonlinear effects.

{\it Hybrid quantum magnetism.--} Our starting point is the model in Eq.\ \eqref{eq:model}, which consists of a set of independent resonators and qubits. The qubit-resonator connections are designed in such a way that each cavity interacts with one or more qubits (and vice versa), but no two cavities and no two qubits interact with each other. Formally, Eq.\ \eqref{eq:model} resembles the physics of Jahn-Teller models\ \cite{porras_jahn_teller} and the Rabi lattice model\ \cite{rabi_lattice_mean_field,JCL_with_CRC,rabi_lattice_mean_field_porras,rabi_lattice_entanglement}. The key difference is that bosons form a flat band: i.e. the bosons cannot, by themselves, transport any long range interaction, and it is only the interplay of spins and bosons that gives rise to a many-body model.

It is clear that this system exhibits quantum magnetism in the dispersive regime  so familiar to circuit-QED. When the qubit-photon coupling is  perturbative $ |g|, \omega_0\ll \omega $, any two qubits connected by a resonator couple through the exchange of virtual photons. This leads to an effective Ising model $
H_{\rm dis} = \frac{\omega_0}{2}\sum_i \sigma^z_i+\sum_{i,j} J_{ij} \sigma^x_i \sigma^x_j $, with  interaction $J_{ij}=-\sum_{r}g_{ir} g_{jr}/\omega$. As outlined in the introduction, dispersive regimes necessarily lead to weak interactions $|J|\ll|g| \ll \omega$.

The important question addressed in this work is whether photon-mediated interactions can reach a regime of ultrastrong quantum magnetism, $|J|\lesssim|g| \approx \omega,\omega_0$. We answer this with a many-body variational ansatz for the ground state of\ \eqref{eq:model} that exploits the polaron transformation to disentangle spins and photons\ \cite{porras_ising}. Namely,%
\ifcheckpagelimits\else%
\begin{equation}
 \ket{\Psi_{\rm GS}(\psi_{\rm spin},\alpha)} = U^{\dagger} \ket{\psi_{\rm spin}} \otimes_i\ket{\alpha_i},
 \label{eq:ansatz}
\end{equation}%
\fi%
where $\ket{\psi_{\rm spin}}$ is an arbitrary many-body state in the spin Hilbert space, photons are parametrized with a product of coherent states, $\ket{\alpha_i}$, and both are entangled by the unitary $U = \ee^{-\ii \sum_i \Theta_i\sigma_i^x/2}$ with $\Theta_i=-\sum_{r} 2\ii g_{ir}( a_r- a^\dagger_r)/{\omega}$. The minimal energy is obtained from the transformed Hamiltonian $H_{\rm eff} = U H U^{\dagger}$,
\ifcheckpagelimits\else%
\begin{equation}
 H_{\rm eff}\! = H_{\rm b}+\! \sum_{i,j} \! J_{ij}\sigma^x_i\sigma^x_j
 + \frac{\omega_0}{2}\! \sum_i\! (\! \sigma^z_i \cos\Theta_i+\sigma^y_i \sin\Theta_i\! )\!.
 \label{Heff}
\end{equation}%
\fi%
This procedure\ \cite{supp_mat} results in a vacuum state for the photons, $\alpha=0$, and the exact ground state of the Ising model, $\ket{\psi_{\rm spin}}=\ket{\psi_{\rm gs}(J_{ij},\tilde{h}_i)}$, under a renormalized transverse field, $\tilde{h}_i=\half \omega_0\ {\rm exp}[-\sum_r2(g_{ir}/\omega)^2]$. Therefore, our ansatz predicts a quantum phase transition in the Ising universality class that depends on the ratio of the spin-spin couplings and the transverse field. Additionally, due to the polaron transformation, the spin ordering extends onto the photons, yielding a hybridized qubit-photon magnetism.

Let us emphasize the following points: {\it (i)} The ferromagnetic ($J_{ij}<0$), or anti-ferromagnetic ($J_{ij}>0$), character of this hybrid magnetism can be designed through the relative sign of the qubit-photon couplings [c.f. Figs.\ \ref{fig:setup}{\bf (a)-(b)}]. {\it (ii)} The geometry of the emerging Ising model is inherited from the qubit-photon connectivity. {\it (iii)} The coupling strengths can reach the ultrastrong-coupling regime, since for $|g|\sim \omega,\omega_0$, one obtains $|J|\sim g\sim \omega,\omega_0$. As announced perviously, our ansatz predicts that {\it the nature of the quantum phase transition is preserved all the way from weak to ultrastrong couplings.}

\begin{figure}[t]
 \ifcheckpagelimits\else%
 \centering
 \includegraphics[width=1\linewidth]{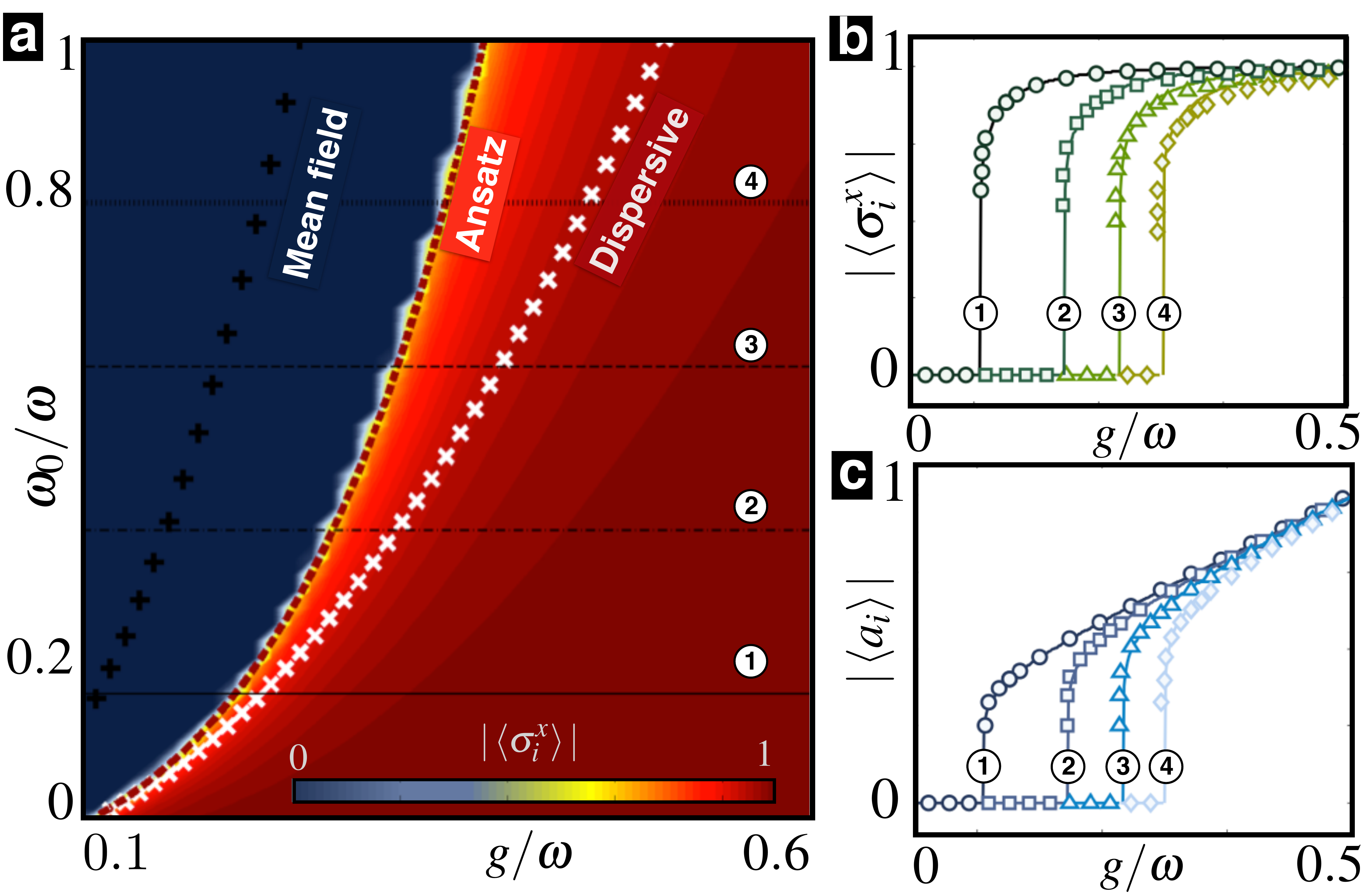}
 \fi
 \caption{{\bf Hybridized spin-photon quantum magnetism:} {\bf (a)} Expectation value of the qubit polarization $|\braket{\sigma^x_i}|$ as a function of $g/\omega$, and $\omega_0/\omega$. The blue region represents a paramagnetic phase, whereas the red region stands for the ordered phase with antiferromagnetic polarisations. We plot the critical lines predicted by mean-field theory (black $\boldsymbol{+}$), dispersive perturbative limit (white $\boldsymbol{\times}$), and our many-body ansatz (red dashed line). The simulation has been done with Eq.~\eqref{eq:model} in a 1D configuration with iTEBD/iMPS methods. {\bf (b, c)} $\braket{\sigma_x}$ and $\braket{a}$ as a function of $g/\omega$ for four different values of the quit frequency displayed in {\bf (a)}. The symbols represent the numerics, and the solid lines stand for the prediction of our ansatz. }
 \label{fig:statics}
\end{figure}
 
To assess the validity of our ansatz, we have simulated numerically Eq.~\eqref{eq:model} for a 1D lattice of interspersed cavities and qubits with antiferromagnetic interactions [Fig.~\ref{fig:setup}{\bf (a)}] using iTEBD/iMPS methods~\cite{itebd}. As shown in Fig.~\ref{fig:statics}{\bf (a)}, the qubits and the cavities spontaneously polarize, acquiring nonzero expectation values of both $|\braket{\sigma^x_i}|$ and $|\braket{a_i}|$ as a function of $\omega_0/\omega,g/\omega$. The critical line predicted by our ansatz, $J=\tilde{h}$ or $\omega_0/\omega=4(g/\omega)^2{\rm exp}{(2g/\omega)^2}$, is shown as a red dashed line in Fig.~\ref{fig:statics}{\bf (a)}. Note that this prediction matches exactly the spontaneous polarization observed numerically, and it  departs from the mean-field and dispersive  predictions\ \cite{supp_mat}. To show that this quantum phase transition belongs to the Ising universality class, we compare the scaling of the numerical magnetization across the critical point with our ansatz's prediction $\braket{\sigma^x_i} = (-1)^i(1-\lambda^2 )^{1/8}\theta(\lambda-1)$, where $\lambda=\tilde{h}/J$ and $\theta(x)$ is the Heaviside step function [Fig.~\ref{fig:statics}{\bf (b)}]. The excellent agreement, even close to the strong coupling regime $g\sim\omega,\omega_0$, shows that the corresponding critical exponent is $\beta=1/8$, in contrast to the mean-field prediction $\beta_{\rm MF}=1/2$. As shown in Fig.~\ref{fig:statics}{\bf (c)}, the cavity polarization also shows excellent agreement with our prediction $\braket{a_i} =(2g/\omega)\langle\sigma_i^x\rangle$, allowing us to extract the critical exponent without measuring the qubits.
 
We note that similar effects have been found for 1D Rabi lattice models using either mean-field theory of the spin-boson lattice model\ \cite{rabi_lattice_mean_field_porras}, valid far away from criticality, or an effective low-energy theory of weakly coupled cavities with ultra-strongly coupled qubits\ \cite{rabi_lattice_mean_field}, which leads to weak quantum magnetism~\cite{comment_rabi_strong}. Fig.~\ref{fig:statics} shows that our model~\eqref{eq:model} leads to strong quantum magnetism, and that our ansatz allows for an accurate prediction  in a wider regime, including the critical region.

{\it Hybridized spin-photon quasiparticles.--} In addition to capturing the static properties of the 1D quantum phase transition, the variational ground-state $\ket{\Omega}=\ket{\psi_{\rm gs}(J_{ij},\tilde{h})}\otimes\ket{{\rm vac}}$ can also be used to approximate the low-energy excitations. Regarding $\ket{\Omega}$ as a filled Fermi sea of Bogoliubov fermions\ \cite{Ising_XY} in a vacuum of photons, the lowest energy excitations are obtained by adding a single delocalized fermion-boson hybrid. Hence,%
\ifcheckpagelimits\else%
\begin{equation}
 \ket{\Psi_{\rm ex}(\{\beta^{\rm f}_q,\beta^{\rm b}_q\})} = U^{\dagger}\sum_q(\beta^{\rm f}_q\gamma_{q+}^{\dagger}+\beta^{{\rm b}}_qa_q^{\dagger})\ket{\Omega},
 \label{eq:ansatz_exc}
\end{equation}%
\fi%
Operators $\gamma_{q+}^{\dagger},a_q^{\dagger}$ create fermionic/bosonic Bloch-wave excitations at quasi-momenta $q\in(0,\pi)$, and $\{\beta^{\rm f}_q,\beta^{\rm b}_q\}$ is the set of complex variational parameters. At low energies, the number of these quasiparticle excitations is small and we can linearise the transformed Hamiltonian~\eqref{Heff} in analogy to the Holstein-Primakoff theory\ \cite{holstein_primakoff}. Then, by means of a variational principle, we obtain a Schr\"odinger-type equation $i\partial_t \boldsymbol{\beta}_q = \mathbb{H}_q \boldsymbol{\beta}_q$ for the variational vector $\boldsymbol{\beta}_q=(\beta^{\rm f}_q,\beta^{\rm b}_q)^t$, where $\mathbb{H}_q$ is a $2\times2$ Hermitian matrix\ \cite{supp_mat}. Its eigenvalues  lead to the  quasiparticle energies $E_{{\rm exc},\pm}(q)=E_{\rm GS}+\epsilon_{{\rm exc},\pm}(q)$ 
\ifcheckpagelimits\else%
\begin{equation}
\epsilon_{{\rm exc},\pm}(q)=\half(\omega_q+\epsilon_q)\pm \half\sqrt{(\omega_q-\epsilon_q)^2+4|\xi_q|^2},
\label{bands}
\end{equation}%
\fi%
where $E_{\rm GS}$ is the ground-state energy of the renormalized Ising model. The energies above the ground-state are specified by the single-particle energies of the Ising model $\epsilon_q=2[(J\cos q+\tilde{h})^2+(J\sin q)^2]^{1/2}$ and by a photonic dispersion $\omega_q=\omega+4\tilde{h}(2g/\omega)^2\sin^2 (q/2)$.

Let us highlight the predictions of our ansatz. At criticality $\lambda=1$, there is a soft mode at $q=\pi$ that becomes gapless, $\epsilon_{{\rm exc},\pm}(\pi+\delta q)\propto (\delta q)^1$, with a dynamical critical exponent of $z=1$. Additionally, the energy gap close to criticality decreases as $\Delta\epsilon_{{\rm exc},\pm}(\pi)\propto |1-\lambda|^1$, which implies that $\nu z=1$, and thus leads to the critical exponent $\nu=1$ in contrast to the mean-field prediction $\nu_{\rm MF}=1/2$. Both predictions are again consistent with the fact that our hybridized spin-photon magnetism lies in the Ising universality class. Below, we confirm the validity of the ansatz~\eqref{eq:ansatz_exc} and the quasiparticle bands~\eqref{bands}, by simulating numerically a possible spectroscopic experiment.

{\it Many-body spectroscopy.--} Spectroscopy is an essential tool for the study of many-body physics. In particular, neutron scattering excels at probing the order and excitations of magnetic materials, because the neutron spin couples to the magnetic structure and has an energy that matches that of the magnetic excitations\ \cite{squires_book}. Unfortunately, this does not generally apply to quantum-optical devices, as either the qubits are pseudo-spins that do not couple to neutrons, or the energy scales are exceedingly different. We propose an alternative spectroscopy to probe the low-energy excitations and recover their momentum and energy ${\bf q}$, $\epsilon_{\rm exc}({\bf q})$.

 \begin{figure}[t]
 \ifcheckpagelimits\else%
 \centering
 \includegraphics[width=1\linewidth]{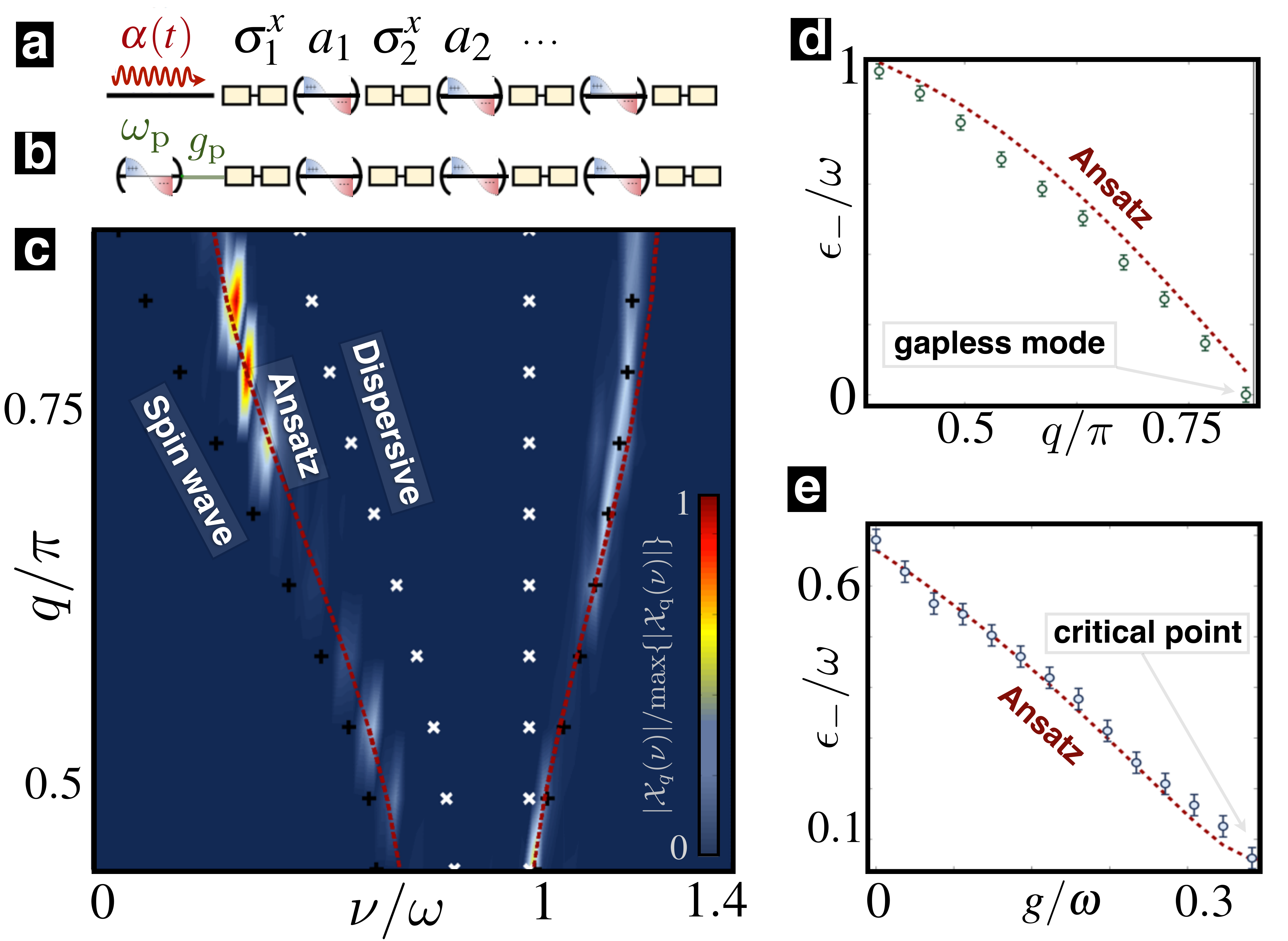}
 \fi
 \caption{{\bf Experimental spectroscopic probes}: {\bf (a)} a coherent drive introduced by a transmission line or {\bf (b)} an equivalent setup with a cavity that hosts a coherent state and is weakly linked to the system, $g_{\rm p}\ll g$. {\bf (c)} Normalized spectroscopic observable $|\mathcal{X}_q(\nu)|$ as a function  of momentum $q/\pi$ and frequency $\nu/\omega$ obtained numerically for the setup in {\bf (b)} with $N=22$ cavities and qubits. The parameters are $g/\omega=0.2$,$\omega_0/\omega=0.69$ for the system, and $g_{\rm p}/\omega=10^{-3}$, $\omega_{\rm p}/\omega=0.1$, and $\alpha_{\rm p}=0.5$. On top of the contour plot, we display the spin-wave prediction for the Gaussian fluctuations over the mean field (black $\boldsymbol{+}$), dispersive perturbative limit (white $\boldsymbol{\times}$), and our many-body ansatz (red dashed line). {\bf (d)} Lower branch excitation energy $\epsilon_-(q)$ as a function of momentum at the critical point, namely $\omega_0/\omega=0.8$ and $g/\omega=0.36$. Green circles correspond to the numerical results, and the red dashed line to the Ansatz. {\bf (e)} Minimum energy gap $\epsilon_-( \pi N/(N+1))$ as a function of $g/\omega$ reaching the critical point for $\omega_0/\omega=0.8$. Blue circles correspond to the numerical results, and the red dashed line to the ansatz.}
 \label{fig:dynamic}
\end{figure}

 The situation that we envision is summarized in Fig.\ \ref{fig:dynamic}{\bf (a)}, where an external but weak drive is used to excite the spin-photon quantum magnet. This results in a set of propagating quasiparticles that induce dynamics in $\braket{x_i(t)}=\braket{a_i+a^\dagger_i},\braket{\sigma^{x}_i(t)}$ that can be experimentally probed ---in particular, the expectation values of $a_i$ can be measured using mobile antennas\ \cite{houck_2013} or qubits that couple to the cavities. Using linear response theory\ \cite{flensberg_book} for a weak driving, one expects that the Fourier transform of $\braket{x_i(t)}$ in momentum and time allows us to recover the Green's function of the model, whose poles will correspond to the quasiparticle energies.
 
To verify this intuition, we have performed numerical simulations using time-dependent MPS\ \cite{garcia_ripoll_2006,verstraete_2008} in a simpler configuration (Fig.\ \ref{fig:dynamic}{\bf (b)}), where the coherent input is provided by a cavity populated with a coherent state and weakly coupled to the quantum magnet ground-state. This initial state is evolved using a Trotter method of third order, computing the observables $\mathbf{X}^{\rm t} = (\braket{x_1(t)},\ldots,\braket{x_N(t)})$. This vector of real numbers is Fourier transformed in position and time
\ifcheckpagelimits\else%
\begin{equation}
 \mathcal{X}_q(\nu) = \frac{1}{T}\sqrt{\frac{2}{N}} \sum_j \int_0^T \mathrm{d}t e^{\ii\nu t} \sin(q j) X_j(t) ,
 \label{ft}
\end{equation}%
\fi%
where $q=\pi/(N+1)\times \mathbb{Z}$ labels the eigenstates of the open boundary conditions problem. 

To model this problem analytically, we have generalized our ansatz~\eqref{eq:ansatz_exc} to account for a boundary cavity with frequency $\omega_{\rm p}$ that couples to the first qubit with strength $g_{\rm p}$~\cite{supp_mat}. In this way, we have proved that the intensity of this signal, $\mathcal{X}_q(\nu)$, is proportional to the zero-temperature retarded cavity-boson Green's function  $G^{\rm b}_{q,{\rm p}}(\nu)=-\ii \int_{-\infty}^{\infty}{\rm d}t \ee^{\ii\nu t}\theta(t)\left\langle [a_q(t),a_{\rm p}^{\dagger}(0)]\right\rangle$, and also to the cavity-fermion Green's function (i.e. $a_q(t)\to\gamma_{q+}(t)$ in the above expression). Using standard techniques, these Green's function can be shown to have poles at the quasiparticle energies $\epsilon_{\rm exc}(q)$, which are shifted and broadened sue to the  cavity-system contact self-energy\ \cite{supp_mat}. However,  when the edge cavity couples very weakly $g_{\rm p}\ll g,\omega,\omega_0$, the contact self-energy is negligible, and the peaks of our observable $\mathcal{X}_q(\nu)$ correspond faithfully to the desired quasiparticle energies.

In Fig.~\ref{fig:dynamic}{\bf (c)}, the color plot shows the numerical results for $|\mathcal{X}_q(\nu)|$. Note the two lines of maxima associated to the two branches of spin-photon excitations: a lower branch that displays a minimum gap around $q=\pi$, and an upper branch centered around $\omega$. The two bands of quasi-particles given by our dynamical ansatz~\eqref{bands} correspond to the red dashed lines, which show a better agreement than spin-wave and dispersive approximations.

{\it Implementation.--} %
A very natural extension of ongoing experiments with 3D cavities would lead to quantum magnetism using high-quality transmons\ \cite{paik,rigetti} that mediate nearby cavities [Fig. \ref{fig:setup}c]. Alternatively, the setup in Fig.\ \ref{fig:setup}b can be implemented using coplanar waveguides and either transmons or galvanically-coupled flux qubits. Coupling strengths are  $g/\omega_0 \sim 0.03$ for transmons\ \cite{majer}, but range from the demonstrated $g/\omega_0 \sim 0.12$  \ \cite{nyemczik} to theoretical limits $g/\omega\sim 1-3$\ \cite{peropadre,bourassa} for flux qubits. In all cases, the critical point and the phase transition are within reach by judiciously choosing the cavity parameters: for the transmon with $\omega_0/2\pi=8$GHz and $g/\omega_0=0.03$, the cavity would be around $300$MHz, while for a flux qubit with $\omega_0=4$GHz and $g/\omega_0=0.2$, the cavity would be around 1.6GHz. Note also that in all the qubits considered so far, the resulting figures are above typical decoherence times $\sim 1\mu$s, and for flux qubits the spin couplings can reach values of GHz, competitive with state-of-the-art Ising interactions in D-Wave setups\ \cite{d_wave}.

Note that, while spin interactions are a recurrent topic in superconducting circuits, the setup proposed in this work represents a non-incremental development that preserves the best features of circuit-QED (strong coupling, fast dynamics, large quality factors) and introduces new detection capabilities. Unlike models\ \cite{ising_direct_couplings} based on direct capacitive\ \cite{capacitive_coupling} or inductive\ \cite{inductive_coupling, squids} qubit-qubit interactions, the presence of cavities as mediators introduces a non-invasive place for local control and measurement of the quantum magnet. Moreover, our model and theory work with strong and ultra-strong spin-spin interactions $|J|\lesssim |g| \approx \omega,\omega_0$. This contrasts with earlier ideas of quantum magnetism in coupled cavities, which either demand  weak cavity-cavity couplings to map the polariton physics onto spin models, $|J|\ll|g| \ll \omega,\omega_0$\ \cite{angelakis_2007}, or are based on dispersive couplings\ \cite{dispersive}, which are weak by definition $|J|\ll |g| \ll |\omega_0-\omega|$.

To perform the measurements, instead of probing the qubits, we suggest using weak links to the individual cavities. These antennae can be permanent, or they can be mobile\ \cite{houck_2013_sm}. They can also measure all cavities, in which case the transform\ \eqref{ft} would be immediate, but it is also possible to recover the dispersion relation from measuring three consecutive cavities and relating their spectra through appropriate linear equations.

Using the same setup it should be possible to do simple transmission-reflection experiments, which are enough to characterize the closing of the gap and extract the static critical exponent. Moreover, relying on incident non-classical states, such as single-photons, opens the door to nonlinear phenomena which are beyond the scope of this work, such as Pauli exclusion and blockade induced quantum gates in the transported excitations\ \cite{gorshkov}.

We finally comment on the possibility of no-go theorems for the existence of  quantum phase transitions, such as those for the Dicke model~\cite{Rzazewski1975,Nataf2010,Viehmann2011}. In these theorems, the physical qubit-cavity coupling induces  an additional renormalization of the cavity frequency $\omega$, which  depends on the coupling strength $g$, and  prevents any phase transition in optical setups~\cite{Nataf2010}. There are two ways to elude a no-go theorem in our case: {\it (i)} we can use flux qubits, where the magnetic dipolar coupling is linear and lacks $A^2$ terms. A state-of-the-art flux qubit can  reach $g_{\rm rel}=g/\omega_0\sim 0.12$~\cite{nyemczik}, implying that the phase transition happens at $\omega=0.21 \omega_0$, a reasonable value. {\it (ii)} We can use charge qubits, where even though capacitive couplings introduce $A^2$ terms, there are enough free parameters to  satisfy the critical relation, as shown in\ \cite{Nataf2010} for the Dicke model, and elaborated in\ \cite{supp_mat} for the present setup. Thus, even if  $g/\omega$ depends non-linearly on the qubit-cavity capacitance, tuning the Josephson energy of the qubit  changes $\omega_0$, such that  the critical point is accesible. In particular, for $g/\omega\in(0,0.4)$ the qubit frequency must be adjusted within  $\omega_0/\omega\in(0,1.21)$. To be more concrete, we note that  renormalization effects depend on the lattice coordination number (e.g. 2 in 1D)\ \cite{supp_mat}, and  can thus never be larger than experimentally observed values with two or more qubits inside a resonator. We  thus use typical  measured values of $\omega,\,g$ and $\omega_0$ to understand whether the phase transition is within reach. For transmons with a coupling $g_{\rm rel}=0.03$  not exceeding their anharmonicity\ \cite{majer},  the critical point is reached for $\omega_0/2\pi=8$GHz and $\omega/2\pi=300$MHz, values within experimental reach and within the regime in which thermal fluctuations would not affect the preparation of the ground state.

{\it Conclusions.--} %
In this work, we have presented a hybrid spin-photon model that implements strong quantum magnetism and Ising-type quantum phase transitions even in regimes of ultrastrong photon-qubit couplings. We have shown how to recover information about the phase transition and critical exponents from static measurements and many-body spectroscopy. We have also  developed a simple yet powerful theory that explains the phase transition and the low-energy excitations.

Finally, and most importantly, our setup allows for very flexible geometric and dynamical design of the interactions. This opens the possibility of: {\it (i)} arbitrary 2D and 3D geometries by a proper design of the qubit-resonator interconnectivity in Eq.~\eqref{eq:model}; {\it (ii)} locally choosing ferromagnetic or antiferromagnetic interactions or introducing disorder, or {\it (iii)} modulating the cavities and qubits to move from Ising to XY models. Hence, this architecture  combines the flexibility of the D-Wave setup for Ising interactions\ \cite{d_wave}, with the speed of resonant ultra-strong qubit-photon couplings. It would also be interesting to combine this setup with the effect of dissipative environments~\cite{debora}.

\ifcheckpagelimits
\else

\acknowledgements
We would like to acknowledge discussions with Diego Porras and David Zueco.  The authors acknowledge support from EU Project PROMISCE, Spanish MINECO Project FIS2012-33022, and CAM regional research consortium QUITEMAD S2009-ESP-1594.

\clearpage

\begin{widetext}

\hypertarget{sm}{\section{Supplemental Material to ``Hybrid quantum magnetism in circuit-QED: from spin-photon waves to many-body spectroscopy''}}

\appendix

In this Supplemental Material, we describe the main steps for the derivations presented in the main text. The contents of this SM are organized as follows:

\begingroup
\hypersetup{linkcolor=black}
\tableofcontents
\endgroup

\section{Ground-state of the 1D spin-boson lattice model}

In this section, we compare the approaches based on mean-field theory, dispersive theory, and our variational ansatz, to predict the ground-state of the spin-boson lattice model in Eq.~\eqref{eq:model}. For concreteness, we restrict to an anti-ferromagnetic one-dimensional chain of $N$ spins and $N$ bosons with all relevant parameters defined below Eq.~\eqref{eq:model},
\begin{equation}
 H = \frac{\omega_0}{2}\sum_{i=1}^N \sigma^z_i + \sum_{i=1}^N  \omega a^\dagger_i a_i
 + \sum_{i=1}^N  g \sigma_i^x (a^\dagger_i -a^\dagger_{i+1} + {\rm H.c.}),
 \label{eq:model_1D}
\end{equation}
 although we note that the approaches can be generalized to other cases straightforwardly.

\subsection{Variational mean-field theory for the spin-boson ground-state}
There is a variety of mean-field (MF) approximations in many-body problems\ \cite{mft}, all of which share the property of neglecting the correlations among the different particles of the system. We focus on variational mean-field theory\ \cite{mft}, which looks for an upper bound of the ground-state energy by minimising over a family of variational product states (i.e. no correlations). For spin systems\ \cite{sw_hcb_sm}, one can propose a mean-field ansatz based on a product of spin coherent states\ \cite{comment_classical}. This procedure can be generalized to the spin-boson model~\eqref{eq:model_1D} by considering the {\it mean-field ansatz as a product state of both spin and bosonic coherent states}. This approach has been used previously for the Rabi lattice model in\ \cite{rabi_lattice_mean_field_porras_sm}, where due to the boson-boson coupling, the correct basis for the coherent states is given by the delocalized collective modes. In our case~\eqref{eq:model_1D}, as the bosons form a localized flat band, we can use the local modes as a basis, such that our mean-field ansatz is
\begin{equation}
\ket{\Psi^{\rm MF}_{\rm GS}(\{\theta_i,\alpha_i\})}=\left(\otimes_i\ee^{-\ii\frac{\theta_i}{2}\sigma_i^y}\ket{-_i}\right)\bigotimes\left(\otimes_i\ee^{\alpha_i a_i^\dagger-\alpha_i^* a_i}\ket{0_i}\right),
\end{equation}
where $\{\theta_i,\alpha_i: i\in\{1,\cdots,N\}\}$ is the set of variational parameters, such that $\theta_i\in\{0,\pi\}$ and $\alpha_i\in\mathbb{C}$. In the definition of the coherent states, we use as reference states $\ket{-_i}=(\ket{\uparrow_i}-\ket{\downarrow_i})/\sqrt{2}$, and the vacuum $\ket{0_i}$. Although this ansatz neglects all possible correlations in the system, it is a good starting point to gain a qualitative understanding.

The variational minimization, $E_{\rm GS}^{\rm MF}={\rm min}\{\bra{\Psi^{\rm MF}_{\rm GS}(\{\theta_i,\alpha_i\})}H\ket{\Psi^{\rm MF}_{\rm GS}(\{\theta_i,\alpha_i\})}\}$, leads to an algebraic system of $3N$ equations. Assuming periodic boundary conditions, and introducing the mean-field coupling $J_{\rm MF}=4g^2/\omega$, and $h=\omega_0/2$, the solution to this system of equations is
\begin{equation}
\begin{split}
\alpha_i&=\{-\alpha_0,+\alpha_0,-\alpha_0,+\alpha_0,\cdots\},\hspace{2ex} \alpha_0=\textstyle{\frac{2g}{\omega}}(1-\lambda_{\rm MF}^2)^{1/2}\theta(1-\lambda_{\rm MF}),\\
\theta_i&=\{\theta_0,\pi-\theta_0,\theta_0,\pi-\theta_0,\cdots\},\hspace{2.1ex}\theta_0=\arccos\left(\textstyle{\frac{\omega}{2g}}\alpha_0\right),
\label{mf_parameters}
\end{split}
\end{equation} where $\lambda_{\rm MF}=h/2J_{\rm MF}$, and $\theta(x)$ is the Heaviside step function (i.e. $\theta(x)=1$ if $x\geq0$, and zero elsewhere). Hence, the mean-field ground-state can be written as follows
\begin{equation}
\ket{\Psi^{\rm MF}_{\rm GS}}=\ket{\theta_0,\pi-\theta_0,\theta_0,\pi-\theta_0,\cdots,\theta_0,\pi-\theta_0}_{\rm spins}\bigotimes\ket{\alpha_0,-\alpha_0,\alpha_0,-\alpha_0,\cdots,\alpha_0,-\alpha_0}_{\rm bosons},
\label{mf_gs}
\end{equation}
which has the energy $E_{\rm GS}^{\rm MF}=-hN\theta(\lambda_{\rm MF}-1)-(h\lambda_{\rm MF}+J_{\rm MF}(1-\lambda_{\rm MF}^2))N\theta(1-\lambda_{\rm MF})$.
We can  identify two phases:

{\it (i) Hybrid paramagnetic phase}: If $h\geq 2J_{\rm MF}$, the variational parameters corresponding to the minimum are $\alpha_i=0$, and $\theta_i=-\pi/2$. The mean-field ground-state is thus 
$
\ket{\Psi^{\rm MF}_{\rm GS}}=\ket{\downarrow,\downarrow,\cdots,\downarrow,\downarrow}_{\rm spins}\bigotimes\ket{0,0,\cdots0,0}_{\rm bosons},$ corresponding to a paramagnetic phase with all spins pointing anti-parallel to the transverse field $h\sigma_i^z$, and the bosonic vacuum. 

{\it (ii) Hybrid anti-ferromagnetic phase:} If $h<2J_{\rm MF}$, we find that $\alpha_0> 0$,  $\theta_0> -\pi/2$, such that the ground-state displays an alternating order $
\ket{\Psi^{\rm MF}_{\rm GS}}=\ket{\theta_0,\pi-\theta_0,\cdots,\theta_0,\pi-\theta_0}_{\rm spins}\bigotimes\ket{\alpha_0,-\alpha_0,\cdots,\alpha_0,-\alpha_0}_{\rm bosons}
$ of both spin and boson polarisations. Here, $\ket{\pm\alpha_0}$ are bosonic coherent states, and $\ket{\theta_0}=\cos(\theta_0/2)\ket{-_i}-\sin(\theta_0/2)\ket{+_i}$, $\ket{\pi-\theta_0}=\sin(\theta_0/2)\ket{-_i}-\cos(\theta_0/2)\ket{+_i}$ are the spin states. Let us note that for $J_{\rm MF}\gg h$, we obtain $\theta_0\approx 0$, and thus recover a hybrid ground-state that resembles a N\'eel-ordered state  $\ket{\Psi^{\rm MF}_{\rm GS}}=\ket{-,+,-,+,\cdots}\bigotimes\ket{\alpha_0,-\alpha_0,\alpha_0,-\alpha_0,\cdots}$.

The MF theory thus predicts a paramagnetic to anti-ferromagnetic phase transition at $h=2J_{\rm MF}$, namely $\omega_0/\omega=16(g/\omega)^2$. This is the parabola represented in Fig.~\ref{fig:statics}{\bf (a)} (black $\boldsymbol{+}$). As customary in MF theories, the exact location of the critical point (or line) differs markedly from the prediction. Yet, MF theory is useful to understand qualitative the  hybrid magnetism that will involve the spontaneous polarization of both spins and bosons. In fact, it predicts  
\begin{equation}
\langle\sigma_i^x\rangle_{\rm MF}=(-1)^i\sqrt{1-\lambda_{\rm MF}^2}\theta(1-\lambda_{\rm MF}),\hspace{3ex} \langle a_i\rangle_{\rm MF}=(-1)^{i+1}\frac{2g}{\omega}\sqrt{1-\lambda_{\rm MF}^2}\theta(1-\lambda_{\rm MF}).
\label{MF_polarizations}
\end{equation}
As outlined in the main text, the magnetic critical exponent would be $\beta_{\rm MF}=1/2$, since $|\langle\sigma_i^x\rangle_{\rm MF}|\sim |1-\lambda_{\rm MF}|^{1/2}$ in the vicinity of the critical point $\lambda_{\rm MF}\to 1^-$.

\subsection{Dispersive theory for the spin-boson ground-state}

Considering the MF results, we have seen that the phase transition depends on the competition of two energy scales, $h$ and $J_{\rm MF}=4g^2/\omega$. The dependence of the latter on the Hamiltonian parameters, namely $g^2/\omega$, suggests that it arises from the spin-boson coupling in second-order perturbation theory $g,\omega_0\ll\omega$. Since the spin-phonon coupling is highly off-resonant $g\ll\omega$ in this regime (i.e. the so-called dispersive regime), it will only excite bosons virtually. Thus, it is possible to trace the bosons out, and obtain a spin dynamics where the {\it spin-spin interactions are mediated by the exchange of such virtual bosons}. The leitmotiv is that, by tracing out the bosons, it might be possible to improve on the MF by keeping the spin correlations in the theory.

This can be formalized by a {\it Schrieffer-Wolff transformation}\ \cite{schrieffer_wolff,atom_photon}, as already used for Jaynes-Cummings and Rabi models on the lattice\ \cite{sw_spin_bosons_ti,sw_spin_bosons_supercond}. For such models, the finite bandwidth of the collective bosonic modes leads to a spin-spin interaction whose range depends on the boson-boson coupling. In the present case~\eqref{eq:model_1D}, as the bosonic modes are localized, the emerging spin-spin interactions will be short-ranged (i.e. nearest neighbor). 
 To obtain the effective model, the first step is to realize that for $g,\omega_0\ll\omega$, the Hilbert space will be clustered in manifolds characterized by the total number of bosons $\mathcal{V}_{N_{\rm T}}={\rm span}\{\ket{n_1,n_2,\cdots,n_N}:\sum_in_i=N_{\rm T}\}$ separated in energies by $\omega$. Then, two distant spins  couple by the virtual exchange of a boson either {\it (i)} going through 
the manifold of one extra boson $\mathcal{V}_{N_{\rm T}+1}$, or {\it (ii)} through a
lower-energy manifold with deficit of one boson $\mathcal{V}_{N_{\rm T}-1}$. Since these virtual processes have opposite detunings, their amplitudes cancel $-g^2(\sigma_i^x-\sigma_{i+1}^x)(\sigma_j^x-\sigma_{j+1}^x)( a_ia^\dagger_j/\omega+ a_i^\dagger a_j/(-\omega))=0$ unless the exchanged boson belongs to the same site $i=j$. In that case, bosonic commutation rules imply that the amplitude is $-(g^2/\omega)(\sigma_i^x-\sigma_{i+1}^x)(\sigma_i^x-\sigma_{i+1}^x)$. Accordingly, this Schrieffer-Wolff transformation leads to the effective Hamiltonian
\begin{equation}
 H_{\rm D}=\sum_i(J\sigma_i^x\sigma_{i+1}^x+h\sigma_i^z)+\sum_i\omega a_i^\dagger a_i,
 \label{disp_H}
 \end{equation}
 where we have introduced the dispersive couplings $J=2g^2/\omega$ and $h=\omega_0/2$. We thus obtain a nearest-neighbor anti-ferromagnetic Ising model in a transverse field, and a collection of uncoupled bosons. Let us highlight that the nearest-neighbor character of the interactions is due to the design of the spin-boson interconnectivity in the lattice model~\eqref{eq:model_1D}. We also note that the MF theory overestimates the spin couplings $J_{\rm MF}=4g^2/\omega$ in comparison to the dispersive limit $J=2g^2/\omega$ valid at $g,\omega_0\ll\omega$.

The effective model~\eqref{disp_H} can be diagonalised exactly\ \cite{pfeuty_ising} by using the so-called {\it Jordan-Wigner transformation}\ \cite{jordan_wigner}, which writes the spin operators in terms of spinless fermions. To set the notation for the following sections, we review this solution here. The first step is to introduce the Jordan-Wigner fermions
\begin{equation}
\sigma_i^z=2f_i^{\dagger}f_i-1,\hspace{2.5ex} \sigma_i^x=f_i^\dagger\ee^{\ii\pi\sum_{j<i}f_j\dagger f_j}+{\rm H.c.},\hspace{2.5ex}\sigma_i^y=-\ii f_i^\dagger\ee^{\ii\pi\sum_{j<i}f_j\dagger f_j}+{\rm H.c.},
\label{jw}
\end{equation}
where $f_i^{\dagger},f_i$ are fermionic creation-annihilation operators. This turns the Hamiltonian~\eqref{disp_H} into a quadratic model of decoupled fermions and bosons $H_{\rm D}=\sum_i(J f_i^\dagger f_{i+1}+Jf_i^\dagger f_{i+1}^\dagger+hf_i^\dagger f_i+{\rm H.c.})+\sum_i\omega a_i^\dagger a_i$, where the bosonic part is already diagonal, and the fermionic one can be diagonalised by going to momentum space $f_j=\sum_q\ee^{\ii q j}f_q/\sqrt{N}$, assuming periodic boundary conditions, and  using  an additional {\it fermionic Bogoliubov transformation}\ \cite{bogoliubov}, which reads
\begin{equation}
\begin{split}
\gamma_{q,+}=u_q^*f_q+v_qf_{-q}^\dagger,\hspace{2ex} \gamma_{q,-}=-v_q^*f_q+u_qf_{-q}^\dagger,
\label{bog}
\end{split}
\end{equation}
where we have introduced $u_q=[\half(1+\Delta_q/\epsilon_q)]^{1/2}$ and $v_q=\ii {\rm sgn(q)}[\half(1-\Delta_q/\epsilon_q)]^{1/2}$ in terms of $\Delta_q=2(J\cos q+h)$, and $\epsilon_q=[\Delta_q^2+(2J\sin q)^2]^{1/2}$. This transformation diagonalizes the dispersive Hamiltonian~\eqref{disp_H}
\begin{equation}
 H_{\rm D}=\sum_{0\leq q\leq \pi}(\epsilon_q\gamma_{q,+}^{\dagger}\gamma_{q,+}-\epsilon_q\gamma_{q,-}^{\dagger}\gamma_{q,-})+\sum_i\omega a_i^\dagger a_i.
 \label{Ising_diag}
 \end{equation}
 From this expression, it is clear that the lowest-energy state will correspond to a product state composed of a Fermi sea with all the negative-energy levels filled and a bosonic vacuum. Hence, 
\begin{equation}
\ket{\Psi^{\rm D}_{\rm GS}}=\left(\otimes_q\gamma_{q,-}^{\dagger}\ket{0}_{\rm fermions}\right)\bigotimes\ket{0}_{\rm bosons},
\label{gs_D}
\end{equation}
whose energy is given by $E_{\rm GS}^{\rm D}=-\sum_q\epsilon_q$.
As announced above, this dispersive theory retains the correlations between the spins, at the expense of working in a regime $g,\omega_0\ll\omega$ where the bosonic ground-state cannot spontaneously polarize. In fact, the equivalent of Eq.~\eqref{MF_polarizations} for the polarizations in the dispersive ground-state~\eqref{gs_D} becomes
\begin{equation}
\langle\sigma_i^x\rangle_{\rm D}=(-1)^i(1-\lambda_{\rm D}^2)^{1/8}\theta(1-\lambda_{\rm D}),\hspace{3ex} \langle a_i\rangle_{\rm D}=0,
\label{D_polarizations}
\end{equation}
where $\lambda_{\rm D}=h/J$. This predicts a critical line at $h=J$, which amounts to the parabola $\omega_0/\omega=4(g/\omega)^2$ displayed in Fig.~\ref{fig:statics}{\bf (a)} (white $\boldsymbol{\times}$). We observe in this figure that, although the prediction is better than the MF theory for $g\ll\omega$, it underestimates the value of the critical point at larger couplings. Moreover, although the theory captures the correct critical exponent $\beta_{\rm D}=1/8$ in the qubit's polarization $|\langle\sigma_i^x\rangle_{\rm D}|\sim |1-\lambda_{\rm D}|^{1/8}$, it also  predicts $\langle a_i\rangle_{\rm D}=0$, which will turn out to be strictly correct only in the limit $g/\omega\to 0 $. Hence, the dispersive theory misses completely the hybrid character of the emerging magnetism of our spin-boson lattice model~\eqref{eq:model_1D}.

\subsection{Variational many-body ansatz for the spin-boson ground-state}
\label{var_ansatz_sm}

In this section, we introduce a many-body ansatz that combines the best of the two previous approximations: {\it (i)} it captures the hybrid character of the magnetic order, and {\it (ii)} it keeps correlations to have a more accurate description of the phase transition. To comply with these requirements, we make use of a {\it Lang-Firsov-type transformation}\ \cite{lang_firsov} for our spin-boson lattice model~\eqref{eq:model_1D}
\begin{equation}
 U = \ee^{-\ii \sum_i \half\Theta_i\sigma_i^x},\hspace{2ex} \Theta_i= -\ii\frac{2 g}{\omega}\left( a_i-a_{i+1}- a^\dagger_i+a^\dagger_{i+1}\right).
 \label{LF}
 \end{equation}
We note that similar transformations have already been used to study Rabi lattice models in the dispersive regime $g,\omega_0\ll\omega$ in the context of trapped ions\ \cite{porras_isin_sm}. In this regime, this {\it Lang-Firsov transformation absorbs all the relevant spin-boson correlations}, and the unitarily-transformed Hamiltonian consists of a set of interacting spins decoupled from the bosons (i.e. analogous to Eq.~\eqref{disp_H}). Our idea is to incorporate this transformation into a many-body ansatz $ \ket{\Psi_{\rm GS}^{\rm A}}=U^{\dagger}\ket{\psi_0}$, where $\ket{\psi_0}$ is some reference state, such that the ansatz is capable of exploring more general regimes (i.e. ultra-strong coupling $g\approx\omega_0,\omega$). Our intuition to choose this reference state is the following: {\it (i)} As the transformation~\eqref{LF} captures most of the spin-boson correlations, we can use a reference state with no additional spin-boson entanglement. {\it (ii)} As the bosons in Eq.~\eqref{eq:model_1D} are not directly coupled, we can use a reference state with no boson-boson correlations. {\it (iii)} As the transformation~\eqref{LF} leads to some effective spin-spin interactions, we have to use a reference state that allows for spin-spin correlations. We thus propose 
\begin{equation}
 \ket{\Psi_{\rm GS}^{\rm A}(\{c_{s_1,s_2,\cdots,s_N},\alpha_i\})} = U^{\dagger} \ket{\psi_{\rm spin}(\{c_{s_1,s_2,\cdots,s_N}\})} \bigotimes\left(\otimes_i\ee^{\alpha_i a_i^\dagger-\alpha_i^* a_i}\ket{0_i}\right),
 \label{eq:ansatz_sm}
\end{equation}
where $\{c_{s_1,s_2,\cdots,s_N}: s_j\in\{\uparrow,\downarrow\}\}$ is a set of $2^N$ complex constants spanning the full spin Hilbert space, and $\{\alpha_i: i\in\{1,\cdots N\}\}$ is a set of $N$ real constants used to define the bosonic coherent states. A variational minimization for~\eqref{eq:ansatz_sm} leads to $E_{\rm GS}={\rm min}\{\bra{\psi_{\rm spin}(\{c_{s_1,s_2,\cdots,s_N}\}),\{\alpha_i\}} H_{\rm eff} \ket{\psi_{\rm spin}(\{c_{s_1,s_2,\cdots,s_N}\}),\{\alpha_i\}} \}$, where the transformed Hamiltonian $H_{\rm eff} = U H U^{\dagger}$ reads
\begin{equation}
 H_{\rm eff}\! = \! \omega\! \! \sum_i a^\dagger_i a_i+\sum_{i} J\sigma^x_i\sigma^x_{i+1}
 + h \sum_i (\sigma^z_i \cos\Theta_i+\sigma^y_i \sin\Theta_i).
 \label{Heff_sm}
\end{equation}
Here, the effective spin-spin interactions have a strength $J=2g^2/\omega$ that coincides with the dispersive calculation~\eqref{disp_H}. Using standard properties of the bosonic coherent states, we find $\bra{\alpha_i}\cos\Theta_i\ket{\alpha_i}=\ee^{-4(g/\omega)^2}$ and $\bra{\alpha_i}\sin\Theta_i\ket{\alpha_i}=0$ since $\alpha_i\in\mathbb{R}$. Accordingly, the variational minimization over the spin and bosonic parameters completely decouples, and we obtain $E_{\rm GS}={\rm min}\{\sum_i\omega|\alpha_i|^2\}+{\rm min}\{\bra{\psi_{\rm spin}(\{c_{s_1,s_2,\cdots,s_N}\})}H_{\rm A} \ket{\psi_{\rm spin}(\{c_{s_1,s_2,\cdots,s_N}\})} \}$ with the following ansatz's spin Hamiltonian
\begin{equation}
 H_{\rm A}\! = \! \sum_{i} \! J\sigma^x_i\sigma^x_{i+1}
 + \tilde{h}\! \sum_i \sigma^z_i ,
 \label{H_A}
\end{equation}
which corresponds to an {\it anti-ferromagnetic Ising model in a renormalized transverse field} $\tilde{h}=h\ee^{-4(g/\omega)^2}$.

The variational problem can be easily solved by following the steps in Eqs.~\eqref{jw}-\eqref{Ising_diag}, making the corresponding substitution $h\to \tilde{h}$ in all expressions. We thus find that $\alpha_i=0$, and the constants $\{c_{s_1,s_2,\cdots,s_N}\}$ correspond to the Bogoliubov fermions in the spin-representation. For concreteness, we rewrite the expressions of the Bogoliubov modes
\begin{equation}
\begin{split}
\tilde{\gamma}_{q,+}=\tilde{u}_q^*f_q+\tilde{v}_qf_{-q}^\dagger,\hspace{2ex} \tilde{\gamma}_{q,-}=-\tilde{v}_q^*f_q+\tilde{u}_qf_{-q}^\dagger,
\label{bog_renorm}
\end{split}
\end{equation}
where we have introduced $\tilde{u}_q=[\half(1+\tilde{\Delta}_q/\tilde{\epsilon}_q)]^{1/2}$, and $\tilde{v}_q=\ii {\rm sgn(q)}[\half(1-\tilde{\Delta}_q/\tilde{\epsilon}_q)]^{1/2}$, in terms of $\tilde{\Delta}_q=2(J\cos q+\tilde{h})$, and $\tilde{\epsilon}_q=[\tilde{\Delta}_q^2+(2J\sin q)^2]^{1/2}$. Hence, our variational ansatz predicts the ground-state
\begin{equation}
\ket{\Psi^{\rm A}_{\rm GS}}=U^{\dagger}\left(\otimes_q\tilde{\gamma}_{q,-}^{\dagger}\ket{0}_{\rm fermions}\bigotimes\ket{0}_{\rm bosons}\right)=:U^{\dagger}\ket{\Omega},
\label{gs_A}
\end{equation}
which has the energy $E_{\rm GS}^{\rm A}=-\sum_q\tilde{\epsilon}_q$.
Let us highlight that, although the result seems equal to the dispersive ground-state~\eqref{gs_D}, there are two crucial differences: {\it (i)} The Bogoliubov fermions in $\ket{\Omega}$ correspond now to a different transverse field $\tilde{h}$, the renormalisation of which  can be understood as the effect of the vacuum fluctuations of the lattice bosons. {\it (ii)} The presence of the Lang-Firsov transformation $U^{\dagger}$ in Eq.~\eqref{gs_A} accounts for the fundamental spin-boson entanglement, and the hybrid nature of the magnetic order. In fact, the analogue of Eq.~\eqref{D_polarizations} for the polarisations within our ansatz is 
\begin{equation}
\langle\sigma_i^x\rangle_{\rm A}=(-1)^i\left(1-\lambda^2\right)^{1/8}\theta(1-\lambda),\hspace{3ex} \langle a_i\rangle_{\rm A}=(-1)^{i+1}\frac{2g}{\omega}\left(1-\lambda^2\right)^{1/8}\theta(1-\lambda),
\label{A_polarizations}
\end{equation}
where $\lambda=\tilde{h}/J$. This predicts a critical line at $\tilde{h}=J$, which amounts to the curve $\omega_0/\omega=4(g/\omega)^2\ee^{4(g/\omega)^2}$ displayed in Fig.~\ref{fig:statics}{\bf (a)} (red dashed line). We observe a remarkable agreement of our prediction with the numerics, even when approaching the ultra-strong coupling regime $g\sim\omega,\omega_0$. Moreover, our many-body ansatz captures the correct critical exponent $\beta=1/8$ in both the qubit's and the boson's polarization. Hence, we can conclude that both spins and bosons display a quantum phase transition whereby their polarization scales according to the Ising universality class. More importantly, the accuracy of our ansatz is preserved all the way, from weak to strong coupling regimes.

\section{On the influence of the $A^2$ term and no-go theorem in circuit-QED}

In this section, we clarify the origin of the model adopted in Eq.~(\ref{eq:model}), and discuss the role of the diamagnetic $A^2$ term. Such a term forbids a superradiant phase transition in the Dicke model describing an ensemble of two-level atoms interacting with a single mode of the electromagnetic field~\cite{Rzazewski1975_sm}. In particular, a sum rule for the dipole oscillator strengths in cavity QED imposes a constraint that is incompatible with the  parameter regime  where a superradiant phase would occur. In contrast, it was recently shown that a superradiant phase transition can take place in a circuit-QED version. Here, superconducting qubits are capacitively coupled to a single resonator, which also leads to the equivalent of the diamagnetic $A^2$ term~\cite{Nataf2010_sm}.

Considering our specific setup of interspersed qubits and  resonators in circuit-QED, we show that this diamagnetic term does not impede the existence of the $\mathbb{Z}_2$ hybrid quantum phase transition either, and that the predictions made with the ansatz in Eq.~(\ref{eq:ansatz}) remain valid if one renormalizes the Hamiltonian parameters adequately.

\subsection{Microscopic derivation of the spin-boson Hamiltonian}

We introduce the Hamiltonian of the lattice of microwave resonators coupled to the charge qubits displayed in Fig.~\ref{fig:setup}{\bf (c)}
\begin{equation}
  H = \sum_i \left[ \frac{q_i^2}{2C} + \frac{1}{2L}\phi_i^2\right]
  + \sum_i \left[\frac{(Q_ i- V_g)^2}{2C_{qb}} - E_J\cos(\Phi_i)\right] 
  + \sum_i \left [  \frac{(q_i - Q_i)^2}{2C_g} + \frac{(q_{i+1} + Q_i)^2}{2C_g} \right ],
\end{equation}
where $q_i$ and $\phi_i$ correspond to the quantized charge and magnetic flux of the resonator with uniform capacitance $C$ and inductance $L$. Similarly, we have the charge $Q_i$ and magnetic flux $\Phi_i$ for the charge qubits, and the Josephson energy term $E_J \cos(\Phi_i) $ with amplitude $E_J$.  Furthermore, $V_g$, $C_g$ describe the effective gate charge and gate capacitance, respectively. Given the Cooper pair charge $2 \text{\ensuremath{e}}$, and the uniform charge energy $E_C$, we may also fix  the capacitive term $C_{qb} = {\text{\ensuremath{e}}}^2/2 E_C$ of the charge qubit.

This Hamiltonian contains three types of terms, which in order of appearance are: {\it (i)} the resonator Hamiltonian, {\it (ii}) the qubit Hamiltonian, and {\it (iii)} the coupling term. In this last term, in addition to the qubit-resonator coupling,  one also finds  a renormalization of the frequency of the resonator, $q_i^2/2C_g$, and of the qubit itself, $Q_i^2/2C_g$. The $q_i^2$ term, when written in terms of Fock operators of the resonator, is the well known $A^2$ term that prevents the Dicke phase transition in a model with many qubits in the same optical cavity. We will now show that in our case, this term does not forbid the spin-boson quantum phase transition.

We will first isolate the harmonic oscillator or bosonic terms. Introducing the coordination number of the lattice, that is the number of qubits that talk to the same cavity (i.e. $z=2$ in Fig.~\ref{fig:setup}{\bf (c)}), this part of the Hamiltonian can be rewritten as
\begin{equation}
  H_{\rm b} = \sum_i \left[ \left(\frac{1}{2C}+z\times\frac{1}{C_g}\right)q_i^2 + \frac{1}{2L}\phi_i^2  \right]\rightarrow H_{\rm b} = \sum_i \tilde{\omega} b^\dagger_i b_i
  \label{eq:renormalized}
\end{equation}
by expressing the charge and flux operators in terms of the bosonic creation-annihilation operators
\begin{equation}
\label{charge_flux}
  q_i =  \ii\sqrt{\frac{{1}}{2\tilde{Z}}}(b_i^\dagger-b_i),\hspace{2ex}   \phi_i =  \sqrt{\frac{\tilde{Z}}{2}}(b_i^\dagger+b_i),
\end{equation}
where we have introduced  the frequency $\tilde{\omega} = 1/\sqrt{L\tilde{C}}$ and impedance $\tilde{Z}=\sqrt{L/\tilde{C}}$, which depend on a renormalized capacitance
\begin{equation}
  \frac{1}{2\tilde{C}} = \frac{1}{2C} + \frac{z}{C_{g}}.
\end{equation}
Note that the renormalization induced on a single resonator is only proportional to the coordination number of the lattice, $z$. This is in clear contrast to the Dicke model, where the frequency of the bosonic mode is renormalized by its coupling with all two-level atoms, and thus depends on the total number of qubits $N$, which may increase to large numbers. Let us also note that the new bosonic operators $b_i,b_i^{\dagger}$ for the renormalized frequency $\tilde{\omega}$ can be expressed in terms of the original operators $a_i,a_i^{\dagger}$ at the bare frequency ${\omega}=1/\sqrt{LC}$ by means of a Bogoliubov transformation, and therefore can be understood as squeezed modes of the resonator. 

Up to an irrelevant constant, the qubit or spin Hamiltonian can be expressed as
\begin{equation}
  H_{\rm s} = \sum_i\left[\frac{(Q_i-\tilde{V}_g)^2}{\tilde{C}_{qb}} - E_J \cos(\Phi_i) \right],
\end{equation}
where we have introduced a renormalized gate charge $\tilde{V}_g=V_g(1+C_{qb}/C_g)$, and qubit capacitance 
\begin{equation}
  \frac{1}{2\tilde{C}_{\rm qb}} = \frac{1}{2C_{\rm qb}} + \frac{z}{C_{g}},
\end{equation}
which is modified due to its coupling to the resonator. By tuning the gate voltage, we can still find a sweet spot where the two lowest energy levels are formed by linear superpositions of states containing a different numbers of Cooper pairs $\ket{\pm}=(\ket{0}\pm \ket{1})/\sqrt{2}$ separated by a qubit energy $\omega_0=-E_{J}$. The corresponding Hamiltonian is $H_{\rm s}=\frac{\omega_0}{2}\sum_i\sigma_i^z$, where $\sigma_i^z=\ket{+_i}\bra{+_i}-\ket{-_i}\bra{-_i}$. We thus see that the qubit-resonator coupling only modifies the condition on the sweet spot, but not the qubit frequency.

Finally, the qubit-resonator or spin-boson coupling is provided by the following term
\begin{equation}
H_{\rm sb}= -\sum_i  \frac{1}{C_g}(q_i-q_{i+1}) Q_i.
\end{equation}
After expressing the resonator charge operator in terms of the Fock operators~\eqref{charge_flux}, and the qubit charge operator in the two-level approximation $Q_i=e(\sigma_i^x-1)$, where  $\sigma_i^x=\ket{+_i}\bra{-_i}+\ket{-_i}\bra{+_i}$, we find that 
\begin{equation}
H_{\rm sb}= \sum_i\ii \tilde{g}(\sigma_i^x-1) (b_i-b_{i+1})+{\rm H.c.}=\sum_i \tilde{g}\sigma_i^x (b_i-b_{i+1})+{\rm H.c.},
\end{equation}
where $\tilde{g}=\ii e/(C_g\sqrt{2\tilde{Z}})$ is the spin-boson coupling strength, and the purely bosonic terms cancel due to the summation. Altogether, 
the  final model, including the renormalizations mentioned above due to the $A^2$ terms, reads
\begin{equation}
  H =\frac{\omega_{0}}{2}\sum_i  \sigma^z_i+ \sum_i  \tilde{\omega} b^\dagger_i b_i + \sum_i  \bigg( \tilde{g} \sigma^x_i (b_i - b_{i+1}) + \mathrm{H.c.}\bigg), 
  \label{modified_H}
\end{equation}
which is  essentially a renormalized version of the form adopted in the manuscript in Eq.~(\ref{eq:model}) with an imaginary spin-boson coupling strength. In the following section, we will show that this slightly modified model still displays a quantum phase transition, and does not suffer from any  no-go theorem.

\subsection{Absence of a No-go theorem and typical setup parameters}

We can repeat the variational procedure of Sec.~\ref{var_ansatz_sm} to determine if the inclusion of the diamagnetic $A^2$ terms ---i.e. the renormalization terms due to $C_g$ leading to the modified Hamiltonian~\eqref{modified_H}--- forbids the possibility of a quantum phase transition. In this case, we need to modify the polaron transformation including the renormalized parameters and the squeezed operators  
\begin{equation}
 U = \ee^{-\ii \sum_i \half\Theta_i\sigma_i^x},\hspace{2ex} \Theta_i= \ii\frac{2 \tilde{g}}{\tilde{\omega}}\left( b_i-b_{i+1}\right)+ {\rm H.c.}
 \label{LF}
 \end{equation}
Additionally, the bosonic part of the variational ansatz~\eqref{eq:ansatz_sm} must now consider purely imaginary variational constants $\alpha_i=\ii|\alpha|$. Performing the variational minimization yields again a renormalized nearest-neighbor Ising model
\begin{equation}
 \tilde{H}_{\rm A}\! = \! \sum_{i} \! \tilde{J}\sigma^x_i\sigma^x_{i+1}
 + \tilde{h}\! \sum_i \sigma^z_i ,
\end{equation}
where $\tilde{J}=2|\tilde{g}|^2/\tilde{\omega}$, and $\tilde{h}=\half\omega_0\ee^{-(2|\tilde{g}|/\tilde{\omega})^2}$. The critical line corresponds to  $\tilde{h}=\tilde{J}$, or equivalently
\begin{equation}
 \omega_0 = \tilde{\omega}\times
 4 \frac{|\tilde g|^2}{\tilde \omega^2}
 e^{4 \frac{|\tilde g|^2}{\tilde \omega^2}},\label{eq:tr-line}
\end{equation}
as explained in the text.

It now remains the question of whether these critical points can be achieved for the microscopic model we considered. As in the circuit-QED setup analyzed by Nataf {\it et al.}~\cite{Nataf2010_sm}, the renormalization of the cavity frequency due to the $A^2$ terms is independent from the qubit frequency itself. In other words, $\omega_0$ is an independent parameter that can be adjusted once the values of $\tilde{g}$ and $\tilde{\omega}$ are known. More precisely, the right-hand side of Eq.\ (\ref{eq:tr-line}) evolves continuously from 0 up to a value which is below $2\tilde{\omega}$ for a realistic range $\tilde{g}/\tilde{\omega} \le 0.4$. This falls well within the dynamic range of $\omega_0$: the value of this frequency can be tuned replacing the junction with a SQUID, so that the total value of $\omega_0$ is upper bounded by the Josephson energy ---normally $E_J \gg \tilde{\omega}$ by one or two orders of magnitude---, and it can be continuously switched down to almost zero.

\begin{figure}[h!]
\includegraphics[width=0.4\linewidth]{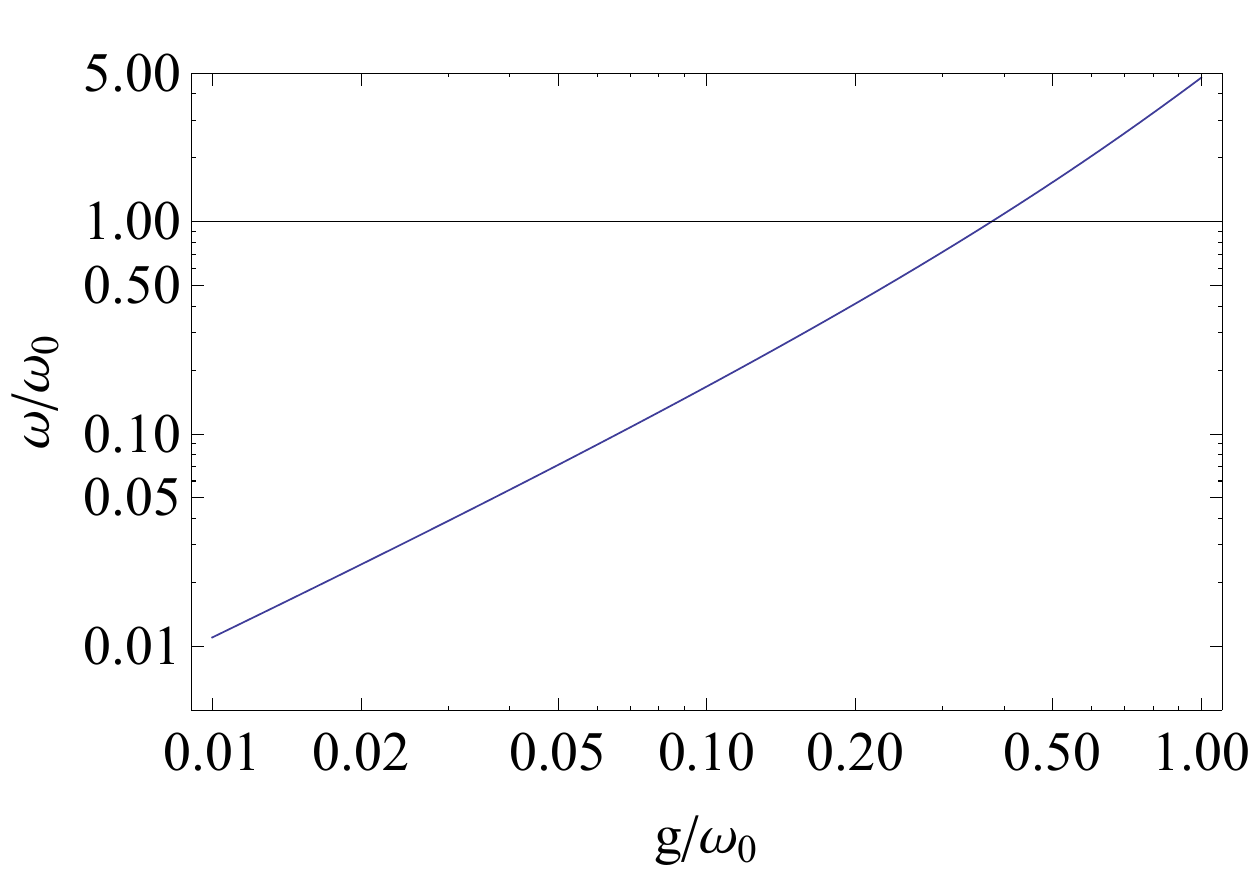}
\caption{Cavity frequency $ \tilde{\omega}/\omega_0$ vs. coupling strength $g_{\rm rel}=\tilde{g}/\omega_0$, all relative to the qubit gap $\omega_0$. This representation is useful because the ratio $\tilde{g}/\omega_0$ must be bounded to preserve the anharmonicity of a transmon qubit.}
\label{fig4}
\end{figure}

The previous is a very particular argument that relies on a charge qubit in our microscopic model. However, using any other qubit ---transmons, phase qubits, flux qubits--- one would arrive to the same quantum Hamiltonian (\ref{modified_H}) with probably larger coupling strengths and better coherence properties. Instead of studying again the capacitances and inductances that would make the phase transition possible and taking into account the renormalization effects, a very simple idea is to just replace $\{\omega_0,\tilde{\omega},\tilde{g}\}$ with typical values of cavity, qubit and coupling energies obtained in different experiments. Note that doing so \textit{we already take into account the renormalization} effects that have been suffered by the qubit and the cavity: as current experiments have been done with one to four qubits, these effects are already larger than those for a setup with coordination number $z=2$ introduced before.

To simplify the analysis we have produced Fig.~\ref{fig4} above, which relates the critical cavity frequency to the relative coupling strength. Let us assume for instance a transmon with coupling strength around $\tilde{g}/\omega_0=0.02$, a value which is still below the anharmonicity and preserves the qubit nature of this circuit. Assuming $\omega_0/2\pi = 8$GHz, the cavity should be around 200 MHz, which is still above significant thermal excitations. Flux qubits are more favorable, as they allow for stronger coupling strengths and do not suffer at all from renormalization effects. In this case, ${g}/\omega_0\sim 0.12$ has already been achieved and ${g}/\omega_0\sim 0.2$ does not seem far away, allowing for a cavity with $0.21\omega_0$ and $0.41\omega_0$, respectively.

\section{Low-energy excitations of the 1D spin-boson lattice model}

In the previous section, we have described three different methods to study the static properties of our spin-boson lattice model~\eqref{eq:model_1D}. In particular, we have observed the supremacy of the many-body ansatz in capturing the hybrid quantum phase transition in a variety of regimes. In this section, we will show that this situation extends to the dynamical properties of the model, since we can build a many-body ansatz that describes the low-energy excitations more accurately than methods based on mean-field (i.e. spin-wave theory), or on the dispersive regime.

\subsection{Spin-wave theory for the low-energy excitations}

For spin models, the low-energy excitations above the mean-field ground-state are known as spin waves (SWs), and there also exists a variety of methods to obtain them\ \cite{auerbach_appendix}. A possible formalism, the so-called {\it Holstein-Primakoff transformation}\ \cite{holstein_primakoff_sm}, describes these spin waves in terms of bosonic operators\ \cite{kubo}, and has also been applied to the Rabi lattice model\ \cite{rabi_lattice_mean_field_porras_sm} by including quantum fluctuations in the bosonic sector of the ground-state.
In this section, we follow this formalism for our model~\eqref{eq:model_1D}.

The first step is to align the spin quantization axis with the direction of the magnetic MF order~\eqref{mf_gs}. This can be accomplished by rotating the spins around the $y$-axis $\sigma_i^x\to\cos\theta_i\tilde{\sigma}_i^x+\sin\theta_i\tilde{\sigma}_i^z$, $\sigma_i^z\to\cos\theta_i\tilde{\sigma}_i^z-\sin\theta_i\tilde{\sigma}_i^z$, where the rotation angle is given by Eq.~\eqref{mf_parameters}. 
We then apply the Holstein-Primakoff transformation
\begin{equation}
\tilde{\sigma}_i^x=(2b_i^{\dagger}b_i-1),\hspace{2.5ex}\tilde{\sigma}_i^y=\ii b_i^{\dagger}\sqrt{1-b_i^{\dagger}b_i}+{\rm H.c.,}\hspace{2.5ex}\tilde{\sigma}_i^z= b_i^{\dagger}\sqrt{1-b_i^{\dagger}b_i}+{\rm H.c.,}
\end{equation}
where $b_i^{\dagger},b_i$ are bosonic operators describing the quantum fluctuations of the spins about the mean-field state, and the corresponding bosonic Hilbert space must be constrained to $\langle b_i^{\dagger}b_i\rangle \leq 1$. For the bosons, the fluctuations are simply $a_i^{\dagger}\to\alpha_i^*+\delta a^{\dagger}_i$, and $a_i\to\alpha_i+\delta a_i$, where $\delta a^{\dagger}_i,\delta a_i$ are also bosonic operators. As customary in linear spin-wave theory, by assuming that these fluctuations are small $\langle b_i^{\dagger}b_i\rangle,\langle \delta a_i^{\dagger}\delta a_i\rangle\ll 1$, we arrive at 
\begin{equation}
H_{\rm SW}=E_{\rm GS}^{\rm MF}+\sum_i\omega \delta a_i^{\dagger}\delta a_i+\sum_i\Delta b_i^{\dagger}b_i+\sum_i \tilde{g}(b_i+b_i^\dagger)(\delta a_i-\delta a_{i+1}+{\rm H.c.}),
\label{sw_hamiltonian}
\end{equation}
where we have introduced the parameters
$
\Delta=4J_{\rm MF}\theta(1-\lambda_{\rm MF}) +2h\theta(\lambda_{\rm MF}-1),$ and $\tilde{g}=-g\lambda_{\rm MF}\theta(1-\lambda_{\rm MF}) -g\theta(\lambda_{\rm MF}-1),
$
and we recall that $\lambda_{\rm MF}=h/2J_{\rm MF}$, and $\theta(x)$ is the Heaviside step function. The above Hamiltonian~\eqref{sw_hamiltonian} describes a collection of coupled harmonic oscillators, such that spin and boson fluctuations are mixed in this hybrid-type of magnetism. 

This Hamiltonian can be diagonalised in momentum representation $b_j=\sum_q\ee^{\ii q j}b_q/\sqrt{N}$, $\delta a_j=\sum_q\ee^{\ii q j}\delta a_q/\sqrt{N}$, where
\begin{equation}
X_q^{a}=\sqrt{\frac{1}{2\omega}}(\delta a_q+\delta a_q^{\dagger}),\hspace{2ex}X_q^{b}=\sqrt{\frac{1}{2\Delta}}( b_q+ b_q^{\dagger}),\hspace{2ex}P_q^{a}=\ii\sqrt{\frac{\omega}{2}}(\delta a^{\dagger}_q-\delta a_q),\hspace{2ex}P_q^{b}=\ii\sqrt{\frac{\Delta}{2}}( b^{\dagger}_q- b_q),\\
\end{equation}
are the spin and boson quadratures.
The spin-wave Hamiltonian can thus be written as $H_{\rm SW}=\half\sum_{q}\boldsymbol{P}_q^{\rm t}\boldsymbol{P}_{-q}+\half\sum_q\boldsymbol{X}_q^{\rm t}\mathbb{K}_{q}\boldsymbol{X}_{-q}$, where $\boldsymbol{P}_{q}=(P_q^{a},P_q^{b})^{\rm t}$, and $\boldsymbol{X}_{q}=(X_q^{a},X_q^{b})^{\rm t}$. The spin-wave dispersion relation, namely the excitation energies above $E_{\rm GS}^{\rm MF}$, is thus given by the square root of the eigenvalues of the following matrix 
\begin{equation}
\mathbb{K}_{q}=\left(\begin{array}{cc}\omega^2 & \tilde{g}_q \\ \tilde{g}^*_q & \Delta^2\end{array}\right),\hspace{2ex}\epsilon^{\rm SW}_{{\rm exc},\pm}(q)=\half\sqrt{(\omega^2+\Delta^2)\pm\sqrt{(\omega^2-\Delta^2)^2+16\omega\Delta|\tilde{g}_q|^2}},
\label{sw_bands}
\end{equation}
where we have introduced $\tilde{g}_q=2\tilde{g}\sqrt{\omega\Delta}(1-\ee^{-\ii q})$. Let us remark that each of the bands~\eqref{sw_bands} describes a hybridized spin-wave branch, such that the excitations contain both spin and bosonic fluctuations about the mean-field ground-state. These energy bands are represented in Fig.~\ref{fig:dynamic}{\bf �} (black $\boldsymbol{+}$), which shows that the SW prediction for the low energy branch is not particularly accurate. In fact, the SW prediction becomes especially deceptive as one approaches the critical region, which is not surprising given the fact that it builds upon a MF approximation that predicts an erroneous critical line (see Fig.~\ref{fig:statics}{\bf (a)}). We have also found by numerical inspection that the SW bands fit the low-energy excitations considerably well far away from criticality.

Additionally, as shown in\ \cite{porras_isin_sm} for the Rabi lattice model, there is an infra-red divergence in both spin and boson fluctuations $\langle b_i^{\dagger}b_i\rangle,\langle \delta a_i^{\dagger}\delta a_i\rangle\to \infty$ as $h\to 2J_{\rm MF}$ in the thermodynamic limit. As this violates the assumptions of validity of the Holstein-Primakoff transformation, namely $\langle b_i^{\dagger}b_i\rangle \leq 1$, one cannot trust the SW predictions close to criticality. For instance, the SW theory allows us to identify the soft mode of the theory , i.e. $\epsilon^{\rm SW}_{{\rm exc},-}(q_{\rm c}=\pi)=0$ at the MF critical point $h=2J_{\rm MF}$, and predict the scalings
\begin{equation}
\epsilon^{\rm SW}_{{\rm exc},-}(\pi)\propto |1-\lambda_{\rm MF}|^{1/2}, \hspace{2ex} \epsilon^{\rm SW}_{{\rm exc},-}(\pi+\delta q)\propto |\delta q|^1.
\end{equation}
This leads to the critical exponents $z_{\rm MF}=1$ and $\nu_{\rm MF}=\half$ by direct comparison with the theory of second-order quantum phase transitions $\epsilon(q_{\rm c}+\delta q)=|\delta q|^{z}$, and $\epsilon(q_{\rm c})\propto |\lambda_{\rm c}-\lambda|^{z\nu}$. As shown in Fig.~\ref{fig:dynamic}{\bf (d)}, this prediction does not fit the numerical results.

\subsection{Dispersive theory for the low-energy excitations}

Let us now turn into the description of the low-energy excitations within the dispersive limit~\eqref{disp_H} of our spin-boson lattice model~\eqref{eq:model_1D}. We recall that the dispersive ground-state $\ket{\Psi^{\rm D}_{\rm GS}}$ in Eq.~\eqref{gs_D} is described by a filled Fermi sea for the spins together with the bosonic vacuum. Therefore, we could write an ansatz for the low-energy excitations as 
\begin{equation}
\ket{\Psi^{\rm D}_{\rm exc}(\{\beta^{\rm f}_q,\beta^{\rm b}_i\})}=\bigg(\sum_q\beta^{\rm f}_q\gamma_{q,+}^{\dagger}+\sum_i\beta_i^{\rm b}a_i^{\dagger}\bigg)\ket{\Psi^{\rm D}_{\rm GS}},
\label{disp_exc}
\end{equation}
where the variational parameters represent the amplitude of creating a fermion over the filled Fermi sea $\beta^{\rm f}_q$, and  a boson over the local vacuum $\beta_i^{\rm b}$. According to Eq.~\eqref{Ising_diag}, we obtain the energy bands above $E_{\rm GS}^{\rm D}$ corresponding to these particle-like excitations
\begin{equation}
\epsilon_{{\rm exc},+}^{\rm D}(q)=\omega,\hspace{2.5ex} \epsilon_{{\rm exc},-}^{\rm D}(q)=2\sqrt{(J\cos q+h)^2+(J\sin q)^2}.
\label{dispersive_bands}
\end{equation}
Here, the higher-branch $\epsilon_{{\rm exc},+}^{\rm D}(q)$ corresponds to localized bosons $\beta^{\rm f}_q=0$, $\beta^{\rm b}_i=\delta_{i,i_0}$, while the 
lower-branch $\epsilon_{{\rm exc},-}^{\rm D}(q_0)$ corresponds to a Bogoliubov fermion with momentum $q_0$, and describes a purely spin-like excitation $\beta^{\rm f}_q=\delta_{q,q_0}$, $\beta^{\rm b}_i=0$. These bands are represented in Fig.~\ref{fig:dynamic}{\bf (c)} (white $\boldsymbol{\times}$), which shows that the lower branch resembles qualitatively the numerical dispersion but it has a large frequency shift. In clear contrast, the dispersive theory misses the behavior of the bosons, as the flat upper bosonic branch is markedly different from the numerical results. Moreover, it misses the hybrid character of the excitations, which are neither purely spin- nor boson-like.

As occurred for the static behavior, the dispersive theory captures correctly the critical exponents. Although it predicts a wrong critical line $h=J$ (Fig.~\ref{fig:statics}{\bf (a)}), it identifies the correct soft mode $q_{\rm c}=\pi$, and the correct scalings
\begin{equation}
\epsilon^{\rm D}_{{\rm exc},-}(\pi)\propto |1-\lambda_{\rm D}|^{1}, \hspace{2ex} \epsilon^{\rm D}_{{\rm exc},-}(\pi+\delta q)\propto |\delta q|^1.
\end{equation}

\subsection{Variational many-body ansatz for the low-energy excitations}

As in the static case, we now introduce a many-body ansatz that combines the best of the two previous approximations: {\it (i)} it captures the hybrid character of the low-energy excitations, and {\it (ii)} it predicts the correct scaling around the critical line. Therefore, we will explore a variational ansatz that is similar to Eq.~\eqref{disp_exc}, but already restricted to particle-like excitations. The main difference is that the excitations are created from the variational ground-state~\eqref{gs_A}, which has already proven to be an adequate description of the spin-boson lattice model~\eqref{eq:model_1D}. Additionally, we also consider delocalised bosonic excitations 
\begin{equation}
\ket{\Psi^{\rm A}_{\rm exc}(\{\beta^{\rm f}_q,\beta^{\rm b}_q\})}=U^{\dagger}\sum_q\left(\beta^{\rm f}_q\gamma_{q,+}^{\dagger}+\beta_q^{\rm b}a_q^{\dagger}\right)\ket{\Omega}=:U^{\dagger}\ket{\Phi^{\rm A}_{\rm exc}(\{\beta^{\rm f}_q,\beta^{\rm b}_q\})}
\label{anstaz_exc}
\end{equation}
where we recall that $\ket{\Omega}$ contains the Fermi sea of the renormalized Ising model in a transverse field, and the boson vacuum. 
We now describe two possible routes to obtain the energy bands of these excitations.

{\it (i) Stationary variational principle:} The excitation energy of the state~\eqref{anstaz_exc} is given by the variational minimization of 
$\epsilon_{\rm exc}^{\rm A}={\rm min}\{\mathcal{E}(\{\boldsymbol{\beta}_q^{\dagger},\boldsymbol{\beta}_q\})/\mathcal{N}(\{\boldsymbol{\beta}_q^{\dagger},\boldsymbol{\beta}_q\})\}$, where $\boldsymbol{\beta}_q=(\beta^{\rm f}_q,\beta^{\rm b}_q)^t$, and we have introduced 
\begin{equation}
\mathcal{E}(\{\boldsymbol{\beta}_q^{\dagger},\boldsymbol{\beta}_q\})=\bra{\Psi^{\rm A}_{\rm exc}(\{\boldsymbol{\beta}_q\})} H-{E}_{\rm GS}^{\rm A}\ket{\Psi^{\rm A}_{\rm exc}(\{\boldsymbol{\beta}_q\})},\hspace{2ex}\mathcal{N}(\{\boldsymbol{\beta}_q^{\dagger},\boldsymbol{\beta}_q\})=
\langle{\Psi^{\rm A}_{\rm exc}(\{\boldsymbol{\beta}_q\})}|\Psi^{\rm A}_{\rm exc}(\{\boldsymbol{\beta}_q\})\rangle,
\label{variational_energy}
\end{equation}
and $E_{\rm GS}^{\rm A}$ is the energy of the ansatz ground-state~\eqref{gs_A}.
Note that, due to the polaron transformation in~\eqref{anstaz_exc}, we can rewrite $\mathcal{E}(\{\boldsymbol{\beta}_q^{\dagger},\boldsymbol{\beta}_q\})=\bra{\Phi^{\rm A}_{\rm exc}(\{\boldsymbol{\beta}_q\})} (H_{\rm eff}-E_{\rm GS}^{\rm A})\ket{\Phi^{\rm A}_{\rm exc}(\{\boldsymbol{\beta}_q\})} $, where the effective Hamiltonian was presented in Eq.~\eqref{Heff_sm}. If we write $\mathcal{N}(\{\boldsymbol{\beta}_q^{\dagger},\boldsymbol{\beta}_q\})=\sum_q\boldsymbol{\beta}_q^{\dagger}\boldsymbol{\beta}_q$  and $\mathcal{E}(\{\boldsymbol{\beta}_q^{\dagger},\boldsymbol{\beta}_q\})=\sum_q\boldsymbol{\beta}_q^{\dagger} \mathbb{H}_q\boldsymbol{\beta}_q$ as quadratic functionals, where  $\mathbb{H}_q$ is a $2\times2$ Hermitian matrix to be computed below, then the minimization under the constraint $\mathcal{N}=1$ yields directly
\begin{equation}
(\mathbb{H}_q-\epsilon_{\rm exc}^{\rm A}(q)\mathbb{I})\boldsymbol{\beta}_q=0, \hspace{1ex}\forall q.
\label{eigenvalues}
\end{equation}
Therefore, the quasiparticle wavefunctions (energies) are given by the eigenstates (eigenvalues) of a $2\times2$ Hermitian matrix.

{\it (ii) Dynamical variational principle:} If we are interested in the dynamics of the system within the subspace defined by the low-energy
excitations, we can employ a principle of stationary action associated to the following Lagrangian
\begin{equation}
\mathcal{L}(\{\boldsymbol{\beta}_q^{\dagger},\boldsymbol{\beta}_q\})=\ii\sum_{q}(\boldsymbol{\beta}_q^{\dagger}\partial_t\boldsymbol{\beta}_q-\partial_t\boldsymbol{\beta}_q^{\dagger}\boldsymbol{\beta}_q)-\mathcal{E}(\{\boldsymbol{\beta}_q^{\dagger},\boldsymbol{\beta}_q\}).
\label{dynamic_variational}
\end{equation}
If the energy is written once more as a quadratic functional, then the corresponding Euler-Lagrange equations lead to a Schr\"{o}dinger-type equation for each momentum $\ii\partial_t\boldsymbol{\beta}_q=\mathbb{H}_q\boldsymbol{\beta}_q$ under a Hamiltonian given by the  $2\times2$ Hermitian matrix $\mathbb{H}_q$. Once more, the quasiparticle wavefunctions (energies) are given by the eigenstates (eigenvalues) in Eq.~\eqref{eigenvalues}. We have introduced this alternative method, as it will turn out to be essential for the section on the many-body spectroscopy.

According to this discussion, both methods lead to the same low-energy quasiparticles, and the remaining task is to obtain the matrix $\mathbb{H}_q$ based on a physically-motivated approximation. In analogy to the linear spin-wave theory, we assume that the fermionic and bosonic fluctuations in~\eqref{anstaz_exc} are small $\langle \gamma_{q,+}^{\dagger}\gamma_{q,+}\rangle,\langle a_q^{\dagger} a_q\rangle\ll 1$. Keeping only quadratic terms in the Hamiltonian, we obtain $H_{\rm eff}\approx H_{\rm b}+H_{\rm f}+H_{\rm bf}$. Here, the purely bosonic part can be written as
\begin{equation}
H_{\rm b}=\sum_{0\leq q\leq\pi}\omega_q a_q^{\dagger}a_q+\omega_{-q} a_{-q}^{\dagger}a_{-q}+\xi_qa_q^{\dagger}a_{-q}^{\dagger}+\xi_q^*a_{-q}a_{q},
\label{boson_linear}
\end{equation}
which corresponds to a band centered around the bare boson frequency $\omega_q=\omega+4\tilde{h}(2g/\omega)^2\cos q$, and with additional two-mode squeezing that couples opposite modes with strength $\xi_q=-\tilde{h}(2g/\omega)^2\ee^{\ii q}$. Let us remark that this term thus corrects the erroneous flat band predicted by the dispersive theory~\eqref{dispersive_bands}. The purely fermionic part corresponds to the Bogoliubov fermions of the renormalized Ising model~\eqref{bog_renorm}, which we rewrite here for convenience
\begin{equation}
H_{\rm f}=\sum_{0\leq q\leq \pi}\tilde{\epsilon}_q\tilde{\gamma}_{q,+}^{\dagger}\tilde{\gamma}_{q,+}-\tilde{\epsilon}_q\tilde{\gamma}_{q,-}^{\dagger}\tilde{\gamma}_{q,-},
\label{fermion_linear}
\end{equation}
where we recall that the fermionic bands have the following expression $\tilde{\epsilon}_q=2[(J\cos q+\tilde{h})^2+(J\sin q)^2]^{1/2}$, and correct for the deficit of spin correlations in the MF~\eqref{mf_gs} and SW theories~\eqref{sw_bands}. Finally, we also obtain a boson-fermion term
\begin{equation}
H_{\rm bf}=\sum_{0\leq q\leq \pi}{g}_{q}(\tilde{\gamma}_{q,+}+\tilde{\gamma}_{q,-})(a_q^{\dagger}-a_{-q})+{\rm H.c.},
\label{boson_fermion_linear}
\end{equation}
where the boson-fermion coupling is given by ${g}_{q}=\tilde{h}(2g/\omega)(1-\ee^{-\ii q})(\tilde{u}_q+\tilde{v}_q^*)$, and the renormalized constants $\tilde{u}_q,\tilde{v}_q$ are defined below Eq.~\eqref{bog_renorm}. We note that this boson-fermion term, together with the Lang-Firsov unitary in~\eqref{anstaz_exc}, are responsible for the hybrid character of the quasiparticles, which are neither purely bosonic nor fermionic in contrast to the dispersive theory~\eqref{dispersive_bands}.

Let us note that in order to arrive at Eq.~\eqref{boson_fermion_linear}, the terms of the Jordan-Wigner string~\eqref{jw} have been dropped. The validity of this approximation primarily  resides on the excellent  agreement of the analytical and numerical predictions. Besides,  the Jordan-Wigner string was also accounted in a sort of Hartree approximation (i.e. the couplings in Eq.~\eqref{boson_fermion_linear} becoming a function of expectation values $\langle \sigma^z\rangle$), but the analytical results did not improve with respect to the previous ones. 

After this derivation, the variational energy~\eqref{variational_energy} associated with this approximate boson-fermion Hamiltonian can be written as a quadratic form $\mathcal{E}(\{\boldsymbol{\beta}_q^{\dagger},\boldsymbol{\beta}_q\})=\sum_q\boldsymbol{\beta}_q^{\dagger} \mathbb{H}_q\boldsymbol{\beta}_q$, such that the quasiparticle energies correspond to 
\begin{equation}
\mathbb{H}_q=\left(\begin{array}{cc}\omega_q & {g}_q \\ {g}^*_q & \epsilon_q\end{array}\right),\hspace{2ex}\epsilon^{\rm A}_{{\rm exc},\pm}(q)=\half{(\omega_q+\tilde{\epsilon}_q)\pm\half\sqrt{(\omega_q-\tilde{\epsilon}_q)^2+4|g_q|^2}}.
\label{anstaz_energies}
\end{equation}
These energy bands are represented in Fig.~\ref{fig:dynamic}{\bf (c)} (red dashed line), which shows that our many-body ansatz outperforms the SW and dispersive predictions. As occurred for the ground-state ansatz (Fig.~\ref{fig:statics}), we have observed a remarkable agreement between the numerics and our quasiparticle ansatz all the way, from a regime of dispersive couplings to the ultra-strong coupling regime $g\sim\omega,\omega_0$. Moreover, our quasiparticle ansatz captures some additional critical exponent according to the Ising universality class
\begin{equation}
\epsilon^{\rm A}_{{\rm exc},-}(\pi)\propto |1-\lambda|^{1}, \hspace{2ex} \epsilon^{\rm A}_{{\rm exc},-}(\pi+\delta q)\propto |\delta q|^1.
\end{equation}
This allows us to obtain the critical exponents, $z=1$ and $\nu=1$,  which agree with the numerical results (Fig.~\ref{fig:dynamic}{\bf (d,e)}).

\section{Many-body spectroscopy of the 1D spin-boson lattice model}

In this section, we discuss a technique to probe the hybrid quantum magnetism of our model. We assume that the system is prepared in the ground-state $\ket{\Psi_{\rm GS}}$ of the Hamiltonian~\eqref{eq:model_1D} for a particular a set of parameters $(\omega,\omega_0,g)$. This can be accomplished by initializing the system in the paramagnetic state $
\ket{\Psi(0)}=\ket{\downarrow,\downarrow,\cdots,\downarrow}_{\rm spins}\bigotimes\ket{0}_{\rm bosons}$ for $\omega,\omega_0\gg g(0)$. Then, by ramping up the coupling $g(t)\to g$ adiabatically, we prepare approximately the ground-state for the particular set of values $(\omega,\omega_0,g)$.

The measurement of  {\it static properties} (see Fig.~\ref{fig:statics}) does not require measuring the spins, as their properties are directly mapped onto the bosons~\eqref{A_polarizations}. For the proposed implementation with superconducting qubits and microwave resonators, it thus suffices to measure a single resonator in the bulk of the chain $\langle a_i\rangle$, in order to infer the static critical properties (i.e. critical line and magnetic exponent $\beta$), which can be done via a fixed or mobile\ \cite{houck_2013_sm} antenna.

The measurement of {\it dynamical properties} (see Fig.~\ref{fig:dynamic}) is not as straightforward, as it requires a protocol that is sensitive to the low-energy quasiparticles and, moreover, gives us the relevant information (i.e. energy bands $\epsilon_{\rm exc}(q)$ and critical exponents $z,\nu$). In condensed-matter systems, this type of information is encoded in the {\it many-body Green's functions}\ \cite{flensberg_book} of the particular model. For instance, the electronic band structure of solids can be inferred from the imaginary part of the single-electron retarded Green's function, which is observed through angle-resolved photoemission spectroscopy (ARPES)\ \cite{arpes}. More relevant to the problem at hand, the energy bands of the spin waves in magnetic systems can be inferred from the two-particle Green's function (i.e. spin-spin correlation functions), which can be observed through neutron scattering\ \cite{squires_book_sm}. As argued in the main text, this method is not applicable to our system, and we must thus find a different spectroscopic protocol that allows us to obtain $\epsilon_{\rm exc}(q)$. 

We describe such a method in this section, which is based on the physical setup depicted in Fig.~\ref{fig:dynamic}{\bf (b)}. Our idea is to probe the many-body spin-boson lattice model through a single cavity that is {\it (i)} initialized in a coherent state, {\it (ii)} perturbatively coupled to the first spin (i.e. non-invasive probe), and {\it (iii)} has a frequency that can be scanned to extract the energy band $\epsilon_{\rm exc} (q)$ from antenna measurements $\{\langle a_i(t)\rangle\}$. The Hamiltonian of the system is $H_{\rm spect}=H_{\rm p}+H_{\rm sp}+H_{\rm s}$, where 
\begin{equation}
H_{\rm p}=\omega_{\rm p} a_{\rm p}^{\dagger}a_{\rm p},\hspace{2ex} H_{\rm sp}=g_{\rm p} (a_{\rm p}^{\dagger}+a_{\rm p})\sigma_1^x,\hspace{2ex}  H_{\rm s} =\half \sum_{i} \omega_0\sigma^z_i + \sum_{i}  \omega a^\dagger_i a_i
 + \sum_{i}  g \sigma_i^x (a^\dagger_i -a^\dagger_{i+1} + {\rm H.c.}),
 \label{sp_hamiltonian}
\end{equation}
are the Hamiltonians of the probe cavity, the system-probe coupling, and the many-body system, respectively. Here, the probe cavity has a frequency $\omega_{\rm p}$, bosonic operators $a_{\rm p}^{\dagger},a_{\rm p}$, and is coupled perturbatively to the first spin of the system $g_{\rm p}\ll g,\omega,\omega_0$. We consider that the initial state of this composite system is $\ket{\Psi(0)}=\ket{\alpha_{\rm p}}\otimes\ket{\Psi_{\rm GS}}$, where $\alpha_{\rm p}\in\mathbb{R}$ is the coherent state amplitude.

\subsection{System-probe retarded Green's function as a spectroscopic observable}

The goal of this subsection is to show that the {\it system-probe retarded Green's functions} for the Hamiltonian~\eqref{sp_hamiltonian}, and the initial state $\ket{\Psi(0)}=\ket{\alpha_{\rm p}}\otimes\ket{\Psi_{\rm GS}}$, can be inferred from time-resolved measurements of the resonators $\{\langle a_i(t)\rangle \}$. The main idea is that a non-invasive probe with $g_{\rm p}\ll g,\omega,\omega_0$, and $\alpha_{\rm p}\ll 1$, will only excite a few low-energy quasiparticles (i.e. $\langle \tilde{\gamma}_{q,+}^{\dagger}\tilde{\gamma}_{q,+}\rangle,\langle a_q^{\dagger} a_q\rangle\ll 1$ according to our many-body ansatz). We will show that the dynamics of $\{\langle a_i(t)\rangle \}$ encodes information about the system-probe retarded Green's functions in frequency-momentum $(\nu,q)$ representation, namely
\begin{equation}
G^{\rm b}_{q,{\rm p}}(\nu)=-\ii \int_{-\infty}^{\infty}{\rm d}t \ee^{\ii\nu t}\theta(t)\left\langle [a_q(t),a_{\rm p}^{\dagger}(0)]\right\rangle,\hspace{2ex} G^{\rm f}_{q,{\rm p}}(\nu)=-\ii \int_{-\infty}^{\infty}{\rm d}t \ee^{\ii\nu t}\theta(t)\left\langle [\tilde{\gamma}_{q,+}(t),a_{\rm p}^{\dagger}(0)]\right\rangle.
\label{sp_gf}
\end{equation}
These Green's functions follow the standard definition\ \cite{flensberg_book_sm}, and describe how an initial bosonic excitation in the probe resonator propagates through the system in the form a bosonic $G^{\rm b}_{q,{\rm p}}(\nu)$, or fermionic $G^{\rm f}_{q,{\rm p}}(\nu)$, excitation with momentum $q$ and energy $\nu$.

To show this connection, let us start by considering the time evolution of $\ket{\Psi_{\rm GS}^{\rm A}}=U^{\dagger}\ket{\Omega}$~\eqref{gs_A}, which can be written as 
\begin{equation}
\ket{\Psi(t)}=\ee^{-\ii(H_{\rm p}+H_{\rm sp}+H_{\rm s})t}\ee^{\alpha_{\rm p} (a_{\rm p}^{\dagger}-a_{\rm p})}\ket{0_p}\otimes U^{\dagger}\ket{\Omega}=U^{\dagger}U_{\rm eff}(t)\ee^{-\ii\int_0^t{\rm d}t'{V}(t')}\ket{0_{\rm p},\Omega},
\end{equation}
where we have introduced the time-evolution operator $U_{\rm eff}(t)=\ee^{-\ii(H_{\rm p}+H_{\rm sp}+H_{\rm eff})t}$ according to the Lang-Firsov-transformed Hamiltonian~\eqref{Heff_sm}, the impulsive perturbation $V(t')=\ii\alpha_{\rm p}(a_{\rm p}^{\dagger}-a_{\rm p})\delta(t')$, and the state $\ket{0_p,\Omega}$ consisting of the total bosonic vacuum and the Fermi sea of Bogoliubov fermions~\eqref{gs_A}.
After introducing $\tilde{V}(t,t')=U_{\rm eff}(t)V(t')U_{\rm eff}^{\dagger}(t)$, we rewrite 
\begin{equation}
\ket{\Psi(t)}=U^{\dagger}\ee^{-\ii\int_0^t{\rm d}t'\tilde{V}(t,t')}U_{\rm eff}(t)\ket{0_{\rm p},\Omega}\approx U^{\dagger}\left(1-\ii\int_0^t{\rm d}t'\tilde{V}(t,t')\right)\ee^{-\ii E_{\rm GS}^{A}t}\ket{0_{\rm p},\Omega},
\end{equation}
where the last step uses the fact that the probe is non-invasive $g_{\rm p}\ll g,\omega,\omega_0$, and $\alpha_{\rm p}\ll 1$, and employs a linear-response-theory-type calculation. We can now readily obtain the 
bosonic observables, which correspond to
\begin{equation}
\langle a_j(t)\rangle= \bra{\Omega}Ua_j(0)U^{\dagger}\ket{\Omega}-\ii\int{\rm d}t' \bra{0_{\rm p}\Omega} [Ua_j(0)U^{\dagger},\tilde{V}(t,t')]\ket{0_{\rm p}\Omega}.
\end{equation}
Using the non-invasive character of the probe once more, the particular expression of the Lang-Firsov transformation~\eqref{LF}, and the static expectation values for the Bogoliubov vacuum, we find that 
\begin{equation}
\langle a_j(t)\rangle=(-1)^{j+1}\frac{2g}{\omega}(1-\lambda^2)^{1/8}\theta(1-\lambda)+\alpha_{\rm p} \big\langle{0_{\rm p}\Omega} \big|\big[a_j(t)+\frac{g}{\omega}\sigma_{j-1}^x(t)-\frac{g}{\omega}\sigma_j^x(t),a_{\rm p}^{\dagger}(0)-a_{\rm p}(0)\big]\big|{0_{\rm p}\Omega}\big\rangle.
\end{equation}
We thus see that the first term of the dynamic observable contains information about the static polarization of the cavities~\eqref{A_polarizations}, while the remaining terms contain two-time correlation functions between the probe and the system. To evaluate these correlation functions, we use our many-body ansatz~\eqref{anstaz_exc}, which assumes that the dynamics takes place within a particular single-quasiparticle subspace. We can then linearise the operators and make the following substitution
\begin{equation}
a_j(t)\to a_j(t)=\frac{1}{\sqrt{N}}\sum_{0\leq q\leq \pi}\ee^{\ii qj}a_q(t)+\ee^{-\ii qj}a_{-q}(t),\hspace{2ex}\sigma_j^x(t)\to\sigma_j^x(t)=\frac{1}{\sqrt{N}}\sum_{0\leq q\leq \pi}(\ee^{\ii qj}(\tilde{u}_q+\tilde{v}_q^*)\tilde{\gamma}_{q,+}(t)+{\rm H.c.}.
\end{equation}
By performing a Fourier transform to momentum space, $\mathcal{A}_k(t)=\sum_j\ee^{-\ii k j}\langle a_j(t)\rangle/\sqrt{N}$ with $k>0$, we find that
\begin{equation}
\mathcal{A}_k(t)= \mathcal{A}_\pi\delta(k-\pi)+\alpha_{\rm p}\delta(k-q)\bigg(\big\langle{0_{\rm p}\Omega} \big|[a_q(t),a_{\rm p}^\dagger(0)]\big|{0_{\rm p}\Omega} \big\rangle+\chi_q\big\langle{0_{\rm p}\Omega} \big|[\tilde{\gamma}_{q,+}(t),a_{\rm p}^\dagger(0)]\big|{0_{\rm p}\Omega} \big\rangle\bigg),
\end{equation}
where we have introduced the magnitude $\mathcal{A}_\pi=-(2g/\omega)(1-\lambda^2)^{1/8}\theta(1-\lambda)$ of the peak that corresponds to the static alternating order in the anti-ferromagnetic phase, and $\chi_q=(g\alpha_{\rm p}/\omega)(\tilde{u}_q+\tilde{v}_q^*)(-1+\ee^{-\ii q})$. Apart from this static contribution, we find other peaks corresponding to the propagation of the initial probe excitation in the form a bosonic or fermionic excitation with a well-defined momentum $q$. 

Finally, by performing a final Fourier transform to frequency space, $\mathcal{A}_k(\nu)=\int_0^{\infty}\ee^{\ii \nu t}\mathcal{A}_k(t)$ with $\nu>0$, we can readily see that the response function contains peaks proportional the aforementioned system-probe Green's functions~\eqref{sp_gf}, namely
\begin{equation}
\mathcal{A}_k(\nu)= \mathcal{A}_\pi\delta(k-\pi)\delta(\nu)+\ii\alpha_{\rm p}\delta(k-q)\left( G^{\rm b}_{q,{\rm p}}(\nu)+\chi_q G^{\rm f}_{q,{\rm p}}(\nu)\right).
\label{response_function}
\end{equation}
Let us finally note that the numerics in Fig.~\ref{fig:dynamic}{\bf �} display the response function $\mathcal{X}_k(\nu)=\mathcal{A}_k(\nu)+{\rm c.c.}$. We thus conclude that, under the approximations of a non-invasive probe and within the quasiparticle ansatz's subspace, the response function gives us directly the system-probe Green's function with momentum and frequency resolution, which in this context describes how well the initial probe excitation is described as bosonic/fermionic quasiparticle with energy $\nu$ and momentum $k$.

\subsection{System-probe retarded Green's function and quasiparticles energies}

In this subsection, we derive a specific expression of the spectroscopic response function~\eqref{response_function} using the dynamical variational technique in Eq.~\eqref{dynamic_variational}. The first step is to extend the ansatz~\eqref{anstaz_exc} to include the probe cavity
\begin{equation}
\ket{\Psi^{\rm A}_{\rm spect}({\boldsymbol{\beta}})}=U^{\dagger}\bigg(\beta_{\rm p}^{\rm b}a_{\rm p}^{\dagger}+\sum_q\beta^{\rm f}_q\gamma_{q,+}^{\dagger}+\sum_q\beta_q^{\rm b}a_q^{\dagger}\bigg)\ket{0_{\rm p},\Omega}, \hspace{2ex}\boldsymbol{\beta}=(\beta_{\rm p}^{\rm b},\beta^{\rm f}_{q_1},\beta^{\rm b}_{q_1},\beta^{\rm f}_{q_2},\beta^{\rm b}_{q_2},\cdots,\beta^{\rm f}_{q_N},\beta^{\rm b}_{q_N})^{\rm t}.
\label{anstaz_spectrosocpy}
\end{equation}
 In order to construct the variational Lagrangian~\eqref{dynamic_variational}, namely 
$
\mathcal{L}(\boldsymbol{\beta}^{\dagger},\boldsymbol{\beta})=\ii(\boldsymbol{\beta}^{\dagger}\partial_t\boldsymbol{\beta}-\partial_t\boldsymbol{\beta}^{\dagger}\boldsymbol{\beta})-\mathcal{E}(\boldsymbol{\beta}^{\dagger},\boldsymbol{\beta})
$,
we need to calculate $\mathcal{E}(\boldsymbol{\beta}^{\dagger},\boldsymbol{\beta})=\bra{\Psi^{\rm A}_{\rm spect}(\boldsymbol{\beta})} H_{\rm spect}-{E}_{\rm GS}^{\rm A}\ket{\Psi^{\rm A}_{\rm spect}(\boldsymbol{\beta})}$ for the spectroscopy Hamiltonian~\eqref{sp_hamiltonian}. In order to do so, we apply the same arguments used to derive the quasiparticle energies~\eqref{anstaz_energies}. We thus linearise the Hamiltonian, which leads to Eqs.~\eqref{boson_linear}-\eqref{boson_fermion_linear} for the system $H_{\rm s}$, $H_{\rm p}=\omega_{\rm p} a_{\rm p}^{\dagger}a_{\rm p}$ for the probe, and the system-probe coupling
\begin{equation}
H_{\rm sp}=\sum_{0\leq q\leq\pi} g_{q,{\rm p}}(\tilde{\gamma}_{q,+}+\tilde{\gamma}_{q,-})(a_{\rm p}+a_{\rm p}^\dagger)+{\rm H.c.},
\end{equation} 
where $g_{{\rm p},q}=g_{\rm p}\ee^{\ii q}(\tilde{u}_q+\tilde{v}_q^*)/\sqrt{N}$. Then, the variational energy can be rewritten as a quadratic form $\mathcal{E}(\boldsymbol{\beta}^{\dagger},\boldsymbol{\beta})=\boldsymbol{\beta}^{\dagger} \mathbb{H}_{\rm spec}\boldsymbol{\beta}$, where 
\begin{equation}
\mathbb{H}_{\rm spec}=\left(\begin{array}{cc}\omega_{\rm p} & \boldsymbol{g}_{{\rm p}} \\ \boldsymbol{g}_{{\rm p}}^{\dagger} & \mathbb{H}_{\rm s}\end{array}\right),\hspace{2ex} \boldsymbol{g}_{{\rm p}}=(g_{{\rm p},q_1},0,g_{{\rm p},q_2},0,\cdots,g_{{\rm p},q_N},0),
\end{equation}
and we have introduced a $2N\times 2N$ Hermitian matrix $\mathbb{H}_{\rm s}=\bigoplus_q\mathbb{H}_q$ given by the direct sum of the $2\times2$ blocks in Eq.~\eqref{anstaz_energies}. The Euler-Lagrange equations lead to a Schr\"{o}dinger-type equation $\ii\partial_t\boldsymbol{\beta}=\mathbb{H}^{\rm spec}_q\boldsymbol{\beta}$, such that $\mathbb{H}^{\rm spec}_q$ plays the role of a single-particle Hamiltonian. As customary in tight-binding transport problems\ \cite{gf_transport}, the retarded Green's function of the spectroscopic system can be obtained by solving a system of equations $[(\nu+\ii\eta)\mathbb{I}-\mathbb{H}_{\rm spec}]\cdot \mathbb{G}(\nu)=\mathbb{I}$, where $\eta\to0^+$. Here, all the retarded Green's functions, including the system-probe Green's functions of interest~\eqref{sp_gf}, are encoded in the following matrix
\begin{equation}
\mathbb{G}(\nu)=\left(\begin{array}{cc}G_{\rm p}(\nu) & \boldsymbol{G}_{{\rm sp}}(\nu) \\ \boldsymbol{G}_{{\rm ps}}(\nu) & \mathbb{G}_{\rm s}(\nu)\end{array}\right),\hspace{2.5ex} \boldsymbol{G}_{{\rm sp}}(\nu)=(G^{\rm f}_{q_1,{\rm p}}(\nu),G^{\rm b}_{q_1,{\rm p}}(\nu),\cdots, G^{\rm f}_{q_N,{\rm p}}(\nu),G^{\rm b}_{q_N,{\rm p}}(\nu)).
\end{equation}
By solving this system of equations, we find that the system-probe Green's functions are expressed as
\begin{equation}
\boldsymbol{G}_{{\rm sp}}(\nu)=\sum_{q,\alpha=\pm}\lambda_q\frac{1}{\nu-\omega_{\rm p}}\boldsymbol{\epsilon}_{q,\alpha}^{\dagger}\frac{1}{\nu-(\epsilon_{\alpha}(q)+{\rm Re}\boldsymbol{\Sigma}(\nu))-\ii{\rm Im\boldsymbol{\Sigma}(\nu)}},
\label{self_energy}
\end{equation}
where we have introduced the constant $\lambda_q=-\boldsymbol{g}_{\rm p}\cdot\boldsymbol{\epsilon}_{q,\alpha}$, the quasiparticle eigenstates $\mathbb{H}_{\rm s}\boldsymbol{\epsilon}_{q,\alpha}={\epsilon_{\alpha}}(q)\boldsymbol{\epsilon}_{q,\alpha}$, and the contact self-energy matrix $\boldsymbol{\Sigma}(\nu)=\boldsymbol{g}_{\rm p}^{\dagger}\boldsymbol{g}_{\rm p}/((\nu-\omega_{\rm p})-\ii\eta)$. This self-energy describes the level shift ${\rm Re}\boldsymbol{\Sigma}(\nu)$ and broadening ${\rm Im}\boldsymbol{\Sigma}(\nu)$ that the quasiparticles experience as a result of their coupling to the probe. As the matrix elements of this self-energy scale with $\boldsymbol{\Sigma}_{nm}\propto g_{\rm p}^2/N$, it is clear that for a non-invasive probe $g_{\rm p}\ll g,\omega,\omega_0$, it becomes negligible $\boldsymbol{\Sigma}(\nu)\to \mathbb{O}$. Hence, {\it $\boldsymbol{G}_{{\rm sp}}(\nu)$ has poles centered and the quasiparticle energies $\epsilon_{\alpha}(q)$}, which translate into the peaks of our response function~\eqref{response_function}. 

We can thus conclude that the Green's function formalism~\eqref{self_energy}, together with the response function~\eqref{response_function}, allows us to predict the main features of the response function displayed in Fig.~\ref{fig:dynamic}{\bf {�}}: {\it (i)} In the anti-ferromagnetic phase, we observe a peak at $\nu=0$ and $k=\pi$, which grows with the system size $\mathcal{A}_\pi\propto\sqrt{N}$. {\it (ii)} We observe a peak centered around the probe cavity frequency $\nu=\omega_{\rm p}$ for all momenta $k>0$. {\it (iii)} We observe a series of peaks centered around the quasiparticle energies $\nu=\epsilon_{\pm}(q)$ for $k=q$, which become better defined as  $g_{\rm p}\ll g,\omega,\omega_0$. In fact, we can calculate the probe back-action through the contact self-energy. All these features are displayed in Fig.~\ref{fig:dynamic}{{\bf (c)}} and additional numerical simulations not shown here. The remarkable agreement between the numerics and our predictions validates both the Green's function approach, and the quasiparticle ansatz~\eqref{anstaz_exc}.

\end{widetext}

\fi


\begin{references}
\bibitem{comment_rabi_strong} By construction in~\cite{rabi_lattice_mean_field}, the cavity-cavity tunneling $J$ of the ultra-strongly coupled  Rabi lattice model  $|g|\geq\omega=\omega_0$  is limited to $J\ll\omega$. An effective low energy theory predicts a Ising model with $H=\sum_iJ^x\Sigma^x_i\Sigma_{i+1}^x+\Delta\Sigma_i^z$, where $J^x=J(g/\omega_0)^2$ and $\Delta=\omega_0{\rm exp}[-(2g/\omega)^2]/2$. Therefore, ultra-strong magnetic couplings $J^x\sim\omega$ can only be achieved  for $g\gg\omega_0$ at the expense of minimising the quantum fluctuations $\Delta\ll\omega$ (i.e. classical magnetism). Otherwise, for $g\gtrsim\omega_0$, one can achieve quantum magnetism  at the expense of being weak $J^x\sim\Delta\ll\omega$. The absence of weak cavity-cavity couplings in our model~\eqref{eq:model} overcomes this problem, and allows for an emerging magnetism that is both quantum and strong.
\end{references}
\end{document}